\documentclass[aps, prb, twocolumn, amssymb, amsmath, showpacs, superscriptaddress]{revtex4-1}
\usepackage[T1]{fontenc}
\usepackage[latin9]{inputenc}
\usepackage{amsmath}
\usepackage{graphicx}
\usepackage{amssymb}
\usepackage{esint}
\usepackage{SIunits}

\begin{document}

\title{Low-energy effective Hamiltonian involving spin-orbit coupling in Silicene and Two-Dimensional Germanium and Tin}

\author{Cheng-Cheng Liu}
\affiliation {Institute of Physics, Chinese Academy of Sciences, Beijing 100190, China}

\author{Hua Jiang}
\email{jianghuaphy@gmail.com}
\affiliation {International Center for Quantum Materials, Peking University}

\author{Yugui Yao}
\email{ygyao@aphy.iphy.ac.cn}
\affiliation {Institute of Physics, Chinese Academy of Sciences, Beijing 100190, China}

\begin{abstract}
Starting from the symmetry aspects and tight-binding method in combination with first-principles calculation, we systematically derive the low-energy effective Hamiltonian involving spin-orbit coupling (SOC) for silicene, which is very general because this Hamiltonian applies to not only the silicene itself but also the low-buckled counterparts of graphene for other group IVA elements Ge and Sn, as well as graphene when the structure returns to the planar geometry. The effective Hamitonian is the analogue to the first graphene quantum spin Hall effect (QSHE) Hamiltonian. Similar to graphene model, the effective SOC in low-buckled geometry opens a gap at Dirac points and establishes QSHE. The effective SOC actually contains first order in the atomic intrinsic SOC strength $\xi_{0}$, while such leading order contribution of SOC vanishes in planar structure.  Therefore, silicene as well as low-buckled counterparts of graphene for other group IVA elements Ge and Sn has much larger gap opened by effective SOC at Dirac points than graphene due to low-buckled geometry and larger atomic intrinsic SOC strength. Further, the more buckled is the structure, the greater is the gap. Therefore, QSHE can be observed in low-buckled Si, Ge, and Sn systems in an experimentally accessible temperature regime. In addition, the Rashba SOC in silicene is intrinsic due to its own low-buckled geometry, which vanishes at Dirac point $K$, while has nonzero value with $\vec{k}$ deviation from the $K$ point. Therefore, the QSHE in silicene is robust against to the intrinsic Rashba SOC.
\end{abstract}


\maketitle

\section{INTRODUCTION}

\begin{figure}
\includegraphics[width=3.5in]{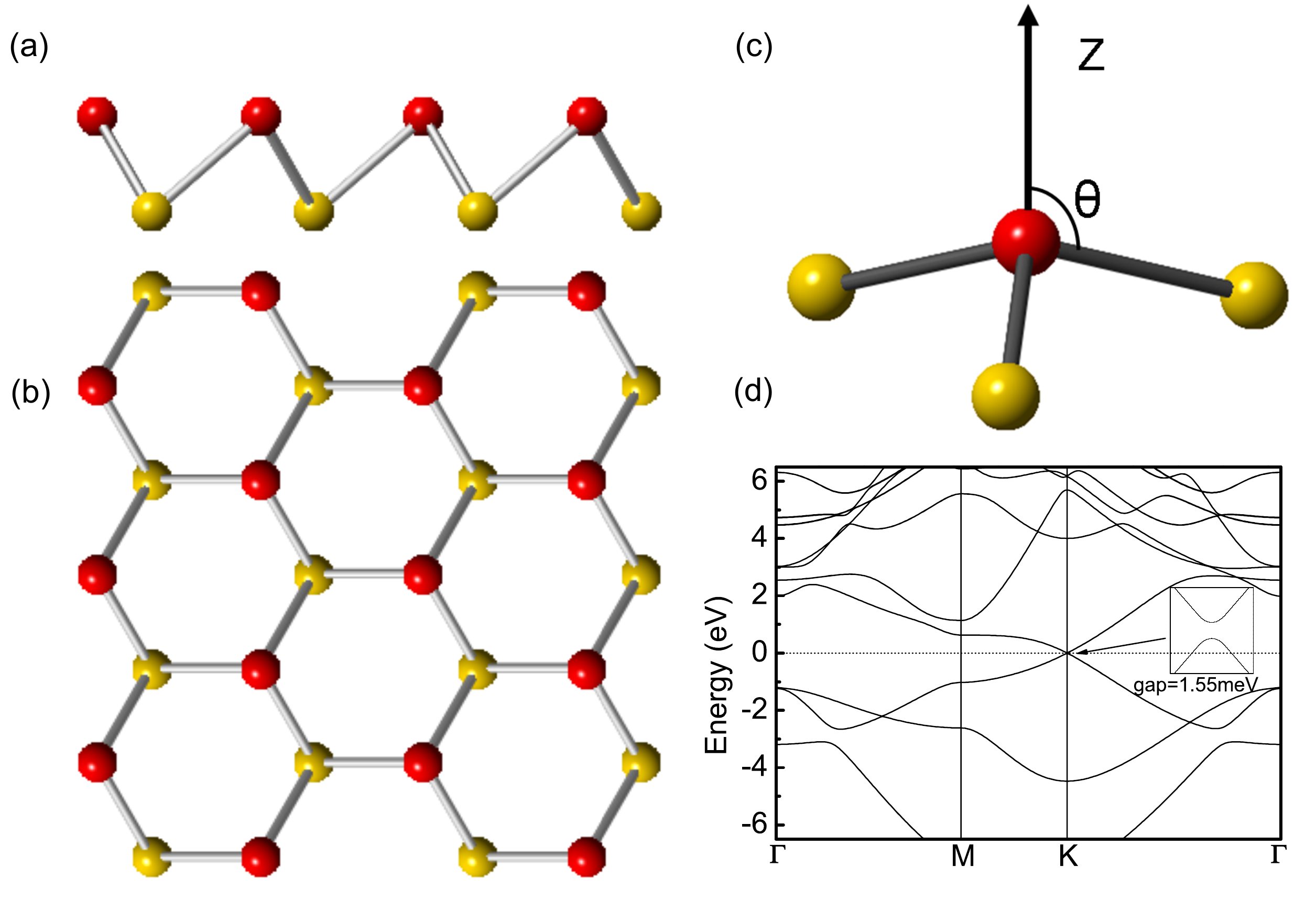}
\caption{(color online). The lattice geometry of low-buckled silicene. (a), (b) The lattice geometry from the side view and top view, respectively. Note that A sublattice (red or gray) and B sublattice (yellow or light gray) are not coplanar. (c) The angle $\theta$ is defined as being between the Si-Si bond and the $z$ direction normal to the plane. (d) The relativistic band structure of low-buckled  silicene. Inset: zooming in the energy dispersion near the $K$ point and the gap induced by SOC.}\label{fig:geometry}
\end{figure}

Silicene, as the counterpart of graphene for silicon, with slightly buckled honeycomb geometry has been synthesized through epitaxial growth~\cite{lalmi_epitaxial_2010}. This novel two-dimensional material has attracted considerable attention both theoretically and experimentally recently, due to exotic electronic structure and promising applications in nanoelectronics as well as compatibility with current silicon-based electronic technology~\cite{
de_padova_evidence_2010,cahangirov_two-_2009,ding_electronic_2009,liu_quantum_2011}.The structure of silicene is shown in Fig.~\ref{fig:geometry}. In the absence of spin-orbit coupling(SOC), the band structure of silicene shows linear energy spectrum crossing at the Fermi level around the Dirac points $K$ and $K^{*}$ of the Hexagonal Brillouin zone~\cite{guzman-verri_electronic_2007,cahangirov_two-_2009,de_padova_evidence_2010,liu_quantum_2011}, which is similar to graphene case.

Quantum spin Hall effect (QSHE), a new quantum state of matter with nontrivial topological property, has garnered great interest in the fields of condensed matter physics and materials science due to its scientific importance as a novel quantum state and the technological applications in spintronics~\cite{hasan_colloquium:_2010,qi_quantum_2010,qi_topological_2010}. This novel electronic state with time reversal invariance is gapped in the bulk and conducts charge and spin in gapless edge states without dissipation at the sample boundaries. The existence of QSHE was first proposed by Kane and Mele in graphene in which SOC opens a band gap at Dirac points~\cite{kane_quantum_2005}. Subsequent works, however, showed that the SOC is rather weak, which is in fact the second order process of the atomic intrinsic spin orbit interaction for graphene, and the QSHE in graphene can occur only at unrealistically low temperature~\cite{yao_spin-orbit_2007,min_intrinsic_2006}. So far, there is only one system, two-dimensional HgTe-CdTe quantum wells, where QSHE is demonstrated~\cite{bernevig_quantum_2006,koenig_quantum_2007}, in spite of some other theoretic suggestions\cite{murakami_quantum_2006,weeks_engineering_2011}. Recently, there is evidence for helical edge modes in inverted InAs-GaSb quantum wells~\cite{liu_quantum_2008} experimentally~\cite{knez_evidence_2011}. Nevertheless, HgTe quantum wells and other systems more or less have serious limitations such as toxicity, difficulty in processing and incompatibility with current silicon-based electronic technology. Therefore, the true realization of QSHE in silicene is very worth while to expect. Silicene and two-dimensional low-buckled honeycomb structures of germanium and tin with QSHE are promising candidates for constructing novel spintronic devices.

Using the first-principles method, we have recently demonstrated that silicene and two-dimensional low-buckled honeycomb structure of germanium can realize the QSHE by exploiting adiabatic continuity and the direct calculation of the $Z_2$ topological invariant~\cite{kane_z2_2005} with a sizable gap opened at the Dirac points due to SOC and the low-buckled structures in our recent Letter~\cite{liu_quantum_2011}. Although the electronic structure, especial linear energy spectrum of silicene, at low energy is similar to that of graphene~\cite{geim_rise_2007,castro_neto_electronic_2009}, the low-buckled geometry makes the derivation of low-energy effective model Hamiltonian not as clear as graphene. Motivated by the fundamental interest associated with QSHE and SOC in silicene, we attempt to give a low-energy effective model Hamiltonian to capture the main physics.

The paper is organized as follows. In Sec. II we briefly describe SOC in silicene from symmetry arguments. Thus, we introduce a next nearest neighbor tight-binding lattice model Hamiltonian to include time reversal invariant spin-orbit interaction. Section III presents the derivation of our low-energy effective model Hamiltonian step-by-step. We investigate in detail the effective spin-orbit interaction including intrinsic Rashba SOC. In Sec. IV, a comparison of gap opened by SOC obtained from between our previous first-principles results and the current tight-binding method is made. As an application of our model Hamiltonian , we also study the counterparts of graphene for other group IVA elements Ge and Sn, which are low-buckled structure according to first-principles calculations. We conclude in Sec. V with a brief discussion and summary.

\section{Lattice model Hamiltonian including spin-orbit coupling in silicene from symmetry aspects}
In general, SOC in Pauli equation can be written as,
\begin{equation}\label{Hsogeneral}
H_{so}=\frac{\hbar}{4m_{0}^{2}c^{2}}\left(\nabla V\times\vec{p}\right)\cdot\vec{\sigma} =-\frac{\hbar}{4m_{0}^{2}c^{2}}\left(\vec{F}\times\vec{p}\right)\cdot\vec{\sigma},
\end{equation}
where $V$ ($\vec{F}$) is potential energy (force), $\vec{p}$ is momentum, $\hbar$ is Plank's constant, $m_{0}$ is the mass of a free electron, c is velocity of light, $\vec{\sigma}$ is the vector of Pauli matrices.

\begin{figure}
\includegraphics[width=3.5in]{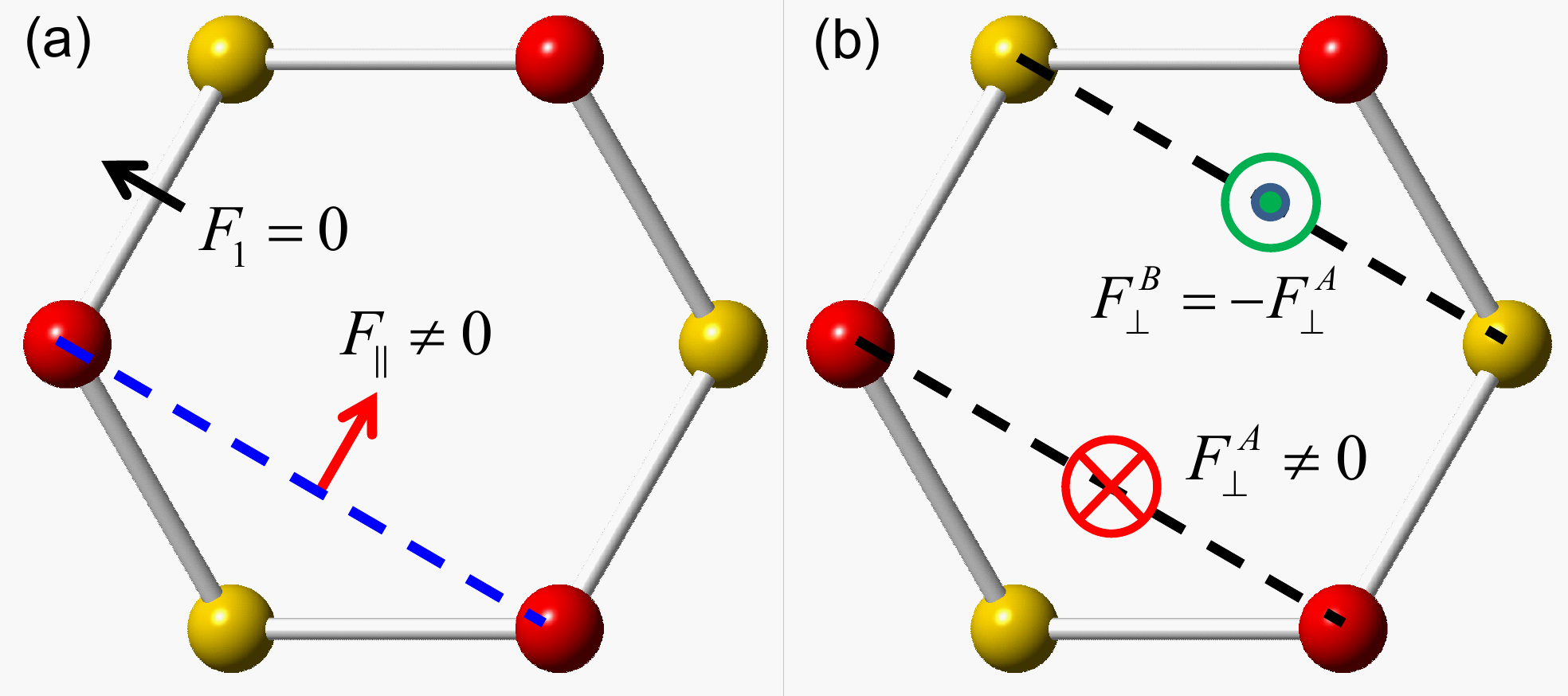}
\caption{(color online). The atomic intrinsic spin-orbit interaction from symmetry aspects. (a) The nearest neighbor force $F_{1}$ vanishes, while the next nearest neighbor force $\vec{F}_{\parallel}$ is non-zero in horizontal plane. (b)  The next nearest neighbor non-zero force $F_{\perp}^{A}$ equals to negative $F_{\perp}^{B}$ in perpendicular direction. }\label{fig:soc}
\end{figure}

For graphene as shown in Fig.~\ref{fig:soc}(a), the nearest neighbor SOC is zero due to its mirror symmetry with respective to an arbitrary bond, while the next nearest neighbor SOC is nonzero. According to symmetry
\begin{equation}
H_{so}=i\gamma_{2}\left(\vec{F}_{\parallel}\times\vec{d}_{ij}\right)\cdot\vec{\sigma}=it_{2}\nu_{ij}\sigma_{z},
\end{equation}
where $\nu_{ij}=\frac{\vec{d}_{i}\times\vec{d}_{j}}{|\vec{d}_{i}\times\vec{d}_{j}|}$, $\gamma_{2}$ and $t_{2}$ are undetermined parameters, $\vec{d}_{i}$ and $\vec{d}_{j}$ are two nearest bonds connecting the next nearest neighbor $\vec{d}_{ij}$.

For silicene, the nearest neighbour SOC is zero, while the next nearest neighbor SOC is nonzero which can be divided into two parts, namely parallel with and perpendicular to the plane, respectively, according to two components of electric field force (see Fig.~\ref{fig:soc}), of which the perpendicular component is due to A sublattice and B sublattice being noncoplanar.

As the first part, the force parallel with the plane is taken into account. This case is similar to the graphene.
\begin{equation}
H_{so1}=i\gamma_{2}\left(\vec{F}_{\parallel}\times\vec{d}_{ij}\right)\cdot\vec{\sigma}\equiv it_{2}\nu_{ij}\sigma_{z}.
\end{equation}
For the second part, the force perpendicular to the plane is taken into account as shown in Fig.~\ref{fig:soc}(b).
\begin{equation}
H_{so2}=i\gamma_{1}\left(\vec{\sigma}\times\vec{d}^{0}_{ij}\right)\cdot F_{\perp}^{A}\vec{e}_{z}\equiv it_{1}\mu_{ij}\left(\vec{\sigma}\times\vec{d}_{ij}^{0}\right)_{z},
\end{equation}
where $\vec{d}_{ij}^{0}=\vec{d}_{ij}/|\vec{d}_{ij}|$, $\gamma_{1}$ and $t_{1}$ are undetermined parameters, $\mu_{ij}=\pm1$ for A site (B site).

Finally, we introduce a second nearest neighbor tight-binding model
\begin{equation}\label{HSmy}
\begin{split}
H=&-t\sum_{\left\langle ij\right\rangle\alpha}c_{i\alpha}^{\dagger}c_{j\alpha}+it_{2}\sum_{\left\langle \left\langle ij\right\rangle \right\rangle \alpha\beta}\nu_{ij}c_{i\alpha}^{\dagger}\sigma_{\alpha\beta}^{z}c_{j\beta}\\
&-it_{1}\sum_{\left\langle \left\langle ij\right\rangle \right\rangle \alpha\beta}\mu_{ij}c_{i\alpha}^{\dagger}\left(\overrightarrow{\sigma}\times\overrightarrow{d_{ij}^{0}}\right)_{\alpha\beta}^{z}c_{j\beta}.
\end{split}
\end{equation}
The first term is the usual nearest neighbor hopping term. The second and third terms are effective SOC and intrinsic Rahsba SOC. The three parameters $t, t_{2}, t_{1}$ are given explicit expression forms in the following derivation by using tight-binding method.

By performing Fourier transformations, we obtain the low-energy effective Hamitoniam around Dirac point K in the basis $\left\{|A\rangle,|B\rangle\right\}\otimes \{ \uparrow, \downarrow \}$
\begin{eqnarray}\label{HeffSmy}
H_{K}^{eff}\approx\left(\begin{array}{cc}
h_{11} & v_{F}(k_{x}+ik_{y})\\
v_{F}(k_{x}-ik_{y}) & -h_{11}
\end{array}\right),
\end{eqnarray}
\begin{equation*}
v_{F}=\frac{\sqrt{3}}{2}at, h_{11}=-3\sqrt{3}t_{2}\sigma_{z}-\frac{3}{2}t_{1}a\left(k_{y}\sigma_{x}-k_{x}\sigma_{y}\right).
\end{equation*}
Around Dirac point $K^*$ in the basis $\left\{|A\rangle,|B\rangle\right\}\otimes \{ \uparrow, \downarrow \}$, we have
\begin{eqnarray}\label{HK*Smy}
H_{K^{*}}^{eff}\approx\left(\begin{array}{cc}
 -h_{11} & v_{F}(k_{x}-ik_{y})\\
 v_{F}(k_{x}+ik_{y}) & h_{11}
\end{array}\right).
\end{eqnarray}
The two effective Hamiltonian Eqs.~(\ref{HeffSmy}) and (\ref{HK*Smy}) should be related by the time-reversal operation.

From the symmetry aspects analysis, we obtain the effective Hamiltonian for silicene that shown as Eqs.~(\ref{HSmy})~(\ref{HeffSmy})~(\ref{HK*Smy}). However, magnitude of the parameter in effective model and microscopic mechanism such as geometry enhanced effective SOC~\cite{liu_quantum_2011} etc are quite unclear. In order to study such effect, we need to construct the effective Hamiltonian from the atomic tight  binding Hamiltonian.

\section{Low-energy effective Hamiltonian from tight-binding theory}

\subsection{Low-energy effective Hamiltonian without SOC}
The outer shell orbitals of silicon, namely $3s$,$3p_{x}$,$3p_{y}$,$3p_{z}$, are naturally taken into account in our analytic calculation. As shown in Fig.~\ref{fig:geometry}, there are A and B two distinct sites in the honeycomb lattice unit cell of silicene. Therefore, in the representation $\{|p_{z}^{A}\rangle,|p_{z}^{B}\rangle,|p_{y}^{A}\rangle,|p_{x}^{A}\rangle,|s^{A}\rangle,|p_{y}^{B}\rangle,|p_{x}^{B}\rangle,|s^{B}\rangle\}$ (For simplicity, the Dirac ket is then omitted over the following context) and at the $K$ point, the total Hamiltonian in Slater Koster frame reads
\begin{eqnarray}\label{H0}
H_{0}&=&\left(\begin{array}{cccccccc}
0 & 0 & 0 & 0 & 0 & V_{3}^{'} & -iV_{3}^{'} & 0\\
0 & 0 & V_{3}^{'} & iV_{3}^{'} & 0 & 0 & 0 & 0\\
0 & V_{3}^{'} & 0 & 0 & 0 & -V_{1}^{'} & -iV_{1}^{'} & V_{2}^{'}\\
0 & -iV_{3}^{'} & 0 & 0 & 0 & -iV_{1}^{'} & V_{1}^{'} & -iV_{2}^{'}\\
0 & 0 & 0 & 0 & \Delta & -V_{2}^{'} & iV_{2}^{'} & 0\\
V_{3}^{'} & 0 & -V_{1}^{'} & iV_{1}^{'} & -V_{2}^{'} & 0 & 0 & 0\\
iV_{3}^{'} & 0 & iV_{1}^{'} & V_{1}^{'} & -iV_{2}^{'} & 0 & 0 & 0\\
0 & 0 & V_{2}^{'} & iV_{2}^{'} & 0 & 0 & 0 & \Delta
\end{array}\right),
\nonumber\\
&&
\end{eqnarray}
where $V_{1}^{'},V_{2}^{'},V_{3}^{'}$ are related to bond parameters ($V_{ss\sigma}$ etc), the detailed derivations are shown in Appendix A. To diagonalize the total Hamiltonian, we take two steps.

Firstly, we perform unitary transformation
\begin{equation}
\begin{split}
& \varphi_{1}^{A}=-\frac{1}{\sqrt{2}}\left(p_{x}^{A}+ip_{y}^{A}\right)=|p_{+}^{A}\rangle, \\
& \varphi_{2}^{B}=\frac{1}{\sqrt{2}}\left(p_{x}^{B}-ip_{y}^{B}\right)=|p_{-}^{B}\rangle, \\
& \varphi_{3}=\frac{1}{\sqrt{2}}\left[-\frac{1}{\sqrt{2}}\left(p_{x}^{A}-ip_{y}^{A}\right)-\frac{1}{\sqrt{2}}\left(p_{x}^{B}+ip_{y}^{B}\right)\right], \\
& \varphi_{4}=\frac{1}{\sqrt{2}}\left[\frac{1}{\sqrt{2}}\left(p_{x}^{A}-ip_{y}^{A}\right)-\frac{1}{\sqrt{2}}\left(p_{x}^{B}+ip_{y}^{B}\right)\right].
\end{split}
\end{equation}
We rewrite the total Hamiltonian in new basis $\left\{p_{z}^{A},s^{A},\varphi_{2}^{B},p_{z}^{B},s^{B},\varphi_{1}^{A},\varphi_{3},\varphi_{4}\right\} $
\begin{eqnarray*}
H_0 \longrightarrow H_{1}=U_{1}^{\dagger}H_{0}U_{1},
\end{eqnarray*}
\begin{eqnarray}
H_{1}=\left(\begin{array}{cccccccc}
0 & 0 & -iV_{3} & 0 & 0 & 0 & 0 & 0\\
0 & \Delta & iV_{2} & 0 & 0 & 0 & 0 & 0\\
iV_{3} & -iV_{2} & 0 & 0 & 0 & 0 & 0 & 0\\
0 & 0 & 0 & 0 & 0 & -iV_{3} & 0 & 0\\
0 & 0 & 0 & 0 & \Delta & -iV_{2} & 0 & 0\\
0 & 0 & 0 & iV_{3} & iV_{2} & 0 & 0 & 0\\
0 & 0 & 0 & 0 & 0 & 0 & V_{1} & 0\\
0 & 0 & 0 & 0 & 0 & 0 & 0 & -V_{1}
\end{array}\right),
\nonumber\\
&&
\end{eqnarray}
where $U_1$ is the unitary matrix that connects the new basis and original basis, $V_{1}=2V_{1}^{'}$, $V_{2}=\sqrt{2}V_{2}^{'}$, $V_{3}=\sqrt{2}V_{3}^{'}$.

Secondly, the new Hamiltonian $H_{1}$ can be separated to three decoupled diagonal blocks, which are named as $H_{A}$, $H_{B}$ and $H_{C}$, respectively. $H_{A}$ reads in the basis $\left\{p_{z}^{A},s^{A},\varphi_{2}^{B}\right\}$
\begin{eqnarray}\label{H_{A}}
H_{A}=\left(\begin{array}{ccc}
 0 & 0 & -iV_{3}\\
 0 & \Delta & iV_{2}\\
 iV_{3} & -iV_{2} & 0
\end{array}\right).
\end{eqnarray}
Its eigenvalues $\varepsilon_{1}$, $\varepsilon_{2}$, $\varepsilon_{3}$ satisfy the eigen-equation
\begin{equation} \label{eigenequation}
E^{3}-\Delta E^{2}-(V_{2}^{2}+V_{3}^{2})E+\Delta V_{3}^{2}=0.
\end{equation}
Since the above equation is cubic equation, the eigenvalues and eigenvectors of $H_A$ can be analytically obtained. We perform unitary transformation
$\left\{ \phi_{1},\phi_{2},\phi_{3}\right\} \equiv\left\{ p_{z}^{A},s^{A},\varphi_{2}^{B}\right\} U_A$,  where
\begin{eqnarray}\label{U_A}
U_{A}=\left(\begin{array}{ccccc}
\frac{1}{\alpha_{1}} &  & \frac{1}{\alpha_{2}} &  & \frac{1}{\alpha_{3}}\\
\\
\frac{V_{2}\varepsilon_{1}}{\alpha_{1}(\Delta-\varepsilon_{1})V_{3}} &  & \frac{V_{2}\varepsilon_{2}}{\alpha_{2}(\Delta-\varepsilon_{2})V_{3}} &  & \frac{V_{2}\varepsilon_{3}}{\alpha_{3}(\Delta-\varepsilon_{3})V_{3}}\\
\\
\frac{i\varepsilon_{1}}{\alpha_{1}V_{3}} &  & \frac{i\varepsilon_{2}}{\alpha_{2}V_{3}} &  & \frac{i\varepsilon_{3}}{\alpha_{3}V_{3}}
\end{array}\right),
\nonumber\\
&&
\end{eqnarray}
with the normalization factors
\begin{equation*}
\alpha_{i}=\sqrt{1+\left[\frac{V_{2}\varepsilon_{i}}{(\Delta-\varepsilon_{i})V_{3}}\right]^{2}+\left(\frac{\varepsilon_{i}}{V_{3}}\right)^{2}} .
\end{equation*}
For simplify, $U_A$ is expressed as $U_A=\left\{u_{ij} \right\}$, where $u_{ij}$ is the matrix element of $U_A$. We rewrite $H_A$ in the new basis $\left\{\phi_{1},\phi_{2},\phi_{3}\right\}$
\begin{eqnarray}
H_A \longrightarrow H_{A}^{'}=U_A^{\dagger}H_{A}U_A=\left(\begin{array}{ccc}
\varepsilon_{1} & 0 & 0\\
0 & \varepsilon_{2} & 0\\
0 & 0 & \varepsilon_{3}
\end{array}\right).
\end{eqnarray}
The above technique in $H_A$ can also apply to the second diagonal block $H_{B}$ which reads in the basis $\left\{p_{z}^{B},s^{B},\varphi_{1}^{A}\right\}$. $H_B$ satisfy the same eigen-equation Eq.~(\ref{eigenequation}). Its eigenvalues $\varepsilon_{4}, \varepsilon_{5}, \varepsilon_{6}$
satisfy:
\begin{equation}
\varepsilon_{4}=\varepsilon_{1},\varepsilon_{5}=\varepsilon_{2},\varepsilon_{6}=\varepsilon_{3}.
\end{equation}
The eigenvectors $U_B$ of $H_B$ are a little different from that of $H_A$.
\begin{eqnarray}\label{U_B}
U_B =\left(
        \begin{array}{ccc}
          u_{11} & u_{12} & u_{13} \\
        -u_{21} & -u_{22} & -u_{23} \\
          u_{31} & u_{32} & u_{33} \\
        \end{array}
      \right),
\end{eqnarray}
where $u_{ij}$ is the matrix element in the unitary matrix $U_A$ as present in Eq.~(\ref{U_A}). We define the unitary transformation $\left\{\phi_{4},\phi_{5},\phi_{6}\right\} =\left\{p_{z}^{B},s^{B},\varphi_{1}^{A}\right\}U_B$. Obviously, $H_B$ is diagonal in new basis.  $H_{C}$ itself is diagonal. We define $\phi_{7}\equiv \varphi_{3}$,$\phi_{8}\equiv \varphi_{4}$.

From Eq.~(\ref{U_A}) to Eq.~(\ref{U_B}), we have found a unitary transformation $U_2$ that connects the original basis  $\left\{p_{z}^{A}, s^{A},\varphi_{2}^{B},p_{z}^{B},s^{B},\varphi_{1}^{A},\varphi_{3},\varphi_{4}\right\}$ and the new basis $ \left\{ \phi_{1},\phi_{4},\phi_{2},\phi_{5},\phi_{3},\phi_{6},\phi_{7},\phi_{8}\right\}$.
Under such unitary transformation, $H_{1}$ will be diagonal.

Combining the above two steps, we finally find the new basis  $ \left\{ \phi_{1},\phi_{4},\phi_{2},\phi_{5},\phi_{3},\phi_{6},\phi_{7},\phi_{8}\right\}$
and the unitary transformation matrix $U=U_1 U_2$ which diagonalize the original Hamiltonian $H_0$. The results were summarized as:
\begin{equation} \label{U}
\begin{split}
& \left\{ \phi_{1},\phi_{4},\phi_{2},\phi_{5},\phi_{3},\phi_{6},\phi_{7},\phi_{8}\right\}   \\
& =\left\{ p_{z}^{A},p_{z}^{B},p_{y}^{A},p_{x}^{A},s^{A},p_{y}^{B},p_{x}^{B},s^{B}\right\} U,  \\
\end{split}
\end{equation}
\begin{eqnarray}\label{H0'}
&&H_0 \longrightarrow H_{0}^{'}=U^{\dagger}H_{0}U , \nonumber\\
&&H_{0}^{'}=\left(\begin{array}{cccccccc}
\varepsilon_{1} & 0 & 0 & 0 & 0 & 0 & 0 & 0\\
0 & \varepsilon_{1} & 0 & 0 & 0 & 0 & 0 & 0\\
0 & 0 & \varepsilon_{2} & 0 & 0 & 0 & 0 & 0\\
0 & 0 & 0 & \varepsilon_{2} & 0 & 0 & 0 & 0\\
0 & 0 & 0 & 0 & \varepsilon_{3} & 0 & 0 & 0\\
0 & 0 & 0 & 0 & 0 & \varepsilon_{3} & 0 & 0\\
0 & 0 & 0 & 0 & 0 & 0 & V_{1} & 0\\
0 & 0 & 0 & 0 & 0 & 0 & 0 & -V_{1}
\end{array}\right).
\end{eqnarray}
So far, the diagonal Hamiltonian has been obtained. Notice that the interesting structure is low-buckled, which means $V_3$ is small due to the angle $\theta$ approaching to $90\degree$. When $V_3$ is small, the three roots of the eigen-equation Eq.~(\ref{eigenequation}) reads
\begin{equation}\label{eigenvalue}
\begin{split}
& \varepsilon_{1}\approx\Delta\frac{V_{3}^{2}}{V_{2}^{2}}, \\
& \varepsilon_{2}\approx\frac{\Delta+\sqrt{\Delta^{2}+4V_{2}^{2}}}{2}, \\
& \varepsilon_{3}\approx\frac{\Delta-\sqrt{\Delta^{2}+4V_{2}^{2}}}{2}.
\end{split}
\end{equation}

Next, we determine the Fermi energy of silicene. Due to its half filling, there are four eigenvalues below the Fermi energy. According to Eqs.~(\ref{H0'}) and (\ref{eigenvalue}), the eigenvalues $\varepsilon_3, V_1$ are below $\varepsilon_{1}$ while the others are above $\varepsilon_{1}$, so the Fermi energy locates around $\varepsilon_1$.  Thus, $\phi_{1}$ and $\phi_{4}$ are low-energy states  which have explicit forms
\begin{equation}\label{phi_1,4}
\begin{split}
& \phi_{1}= u_{11} p_z^A +  u_{21} s ^A +  u_{31} [\frac{1}{\sqrt{2}}(p_x^B-i p_y ^B)],    \\
& \phi_{4}= u_{11} p_z^B - u_{21} s ^B +  u_{31} [-\frac{1}{\sqrt{2}}(p_x^A+i p_y ^A)].
\end{split}
\end{equation}

In order to study the low-energy physics near the Dirac $K$ point,  we perform the small $\vec{k}$  expansion around $K$ by $\vec{k}\rightarrow\vec{k}+K$ and project the Hamiltonian to  the representation $\left\{\phi_{1},\phi_{4}\right\}$. We keep the first order term of $\vec{k}$
\begin{eqnarray}\label{HK}
H_{K}=\varepsilon_{1}I_2+\left(\begin{array}{cc}
 0 & v_{F}k_{+}\\
 v_{F}k_{-} & 0
\end{array}\right),
\end{eqnarray}
with the Fermi velocity $v_{F}$
\begin{eqnarray*}
& v_{F}=\frac{-\sqrt{3}a}{2}[u_{11}^{2}\left(V_{pp\pi}\sin^{2}\theta+V_{pp\sigma}\cos^{2}\theta\right)-u_{21}^{2}V_{ss\sigma} \\
& +2u_{11}u_{21}\cos\theta V_{sp\sigma}-\frac{1}{2}|u_{31}|^{2}\sin^{2}\theta\left(V_{pp\sigma}-V_{pp\pi}\right)],
\end{eqnarray*}
\begin{equation}\label{vf}
k_{+}=k_{x}+ik_{y},k_{-}=k_{x}-ik_{y},
\end{equation}
where $a$ is the lattice constant, $\theta$ is the angle between the Si-Si bond and $z$ direction. Notice we have let $\hbar=1$. So when we calculate Fermi velocity $v_{F}$, $\hbar$ should be considered.

Eq.~(\ref{HK}) and Eq.~(\ref{vf}) are the final low-energy effective Hamiltonian without SOC. The two important results can obtained from these two equations. Firstly, similar to graphene, the low-buckled silicene remains gapless with linear dispersion. Secondly, $ v_{F}$ here is original from all parameters $V_{pp\pi}, V_{pp\sigma},V_{ss\sigma}, V_{sp\sigma}$, while the Fermi velocity $v_{F}$ is only determined by parameter $V_{pp\pi}$ in graphene (when $\theta=\frac{\pi}{2}$, $ v_{F}=-\frac{\sqrt{3}}{2} V_{pp\pi}a$ ).

\subsection{Low-energy effective Hamiltonian with SOC}
The form of SOC Hamiltonian $H_{so}$ is given in the representation $\left\{ p_{z}^{A},p_{z}^{B},p_{y}^{A},p_{x}^{A},s^{A},p_{y}^{B},p_{x}^{B},s^{B}\right\} \otimes \{\uparrow, \downarrow \}$ (Appendix B). We know that Hamiltonian without SOC in the basis set $\left\{ \phi_{1},\phi_{4},\phi_{2},\phi_{5},\phi_{3},\phi_{6},\phi_{7},\phi_{8}\right\} \otimes \{\uparrow, \downarrow \}$ is diagonal from the above depiction. The two representations are related by unitary transformation (Eq.~\eqref{U})
\begin{equation}
U_{so}=U\otimes I_{2},
\end{equation}
where $I_{2}$ is $2 \times 2$ identity matrix for the spin degree of freedom. In the  representation of $\left\{ \phi_{1},\phi_{4},\phi_{2},\phi_{5},\phi_{3},\phi_{6},\phi_{7},\phi_{8}\right\}   \otimes \{\uparrow, \downarrow \}$, SOC Hamiltonian $H_{so}^{'}$ and total Hamiltonian $H^{'}$ read
\begin{equation}\label{Hso'matrix}
\begin{split}
&H_{so}^{'} \longrightarrow H_{so}^{'} = U_{so}^{\dagger}H_{so}U_{so}, \\
&H^{'}      \longrightarrow H^{'}= H_0^{'} \otimes I_2 + H_{so}^{'}.
\end{split}
\end{equation}
The first $4\times4$ diagonal block in SOC Hamiltonian $H_{so}^{'}$ is no other than the first order SOC, which reads at the Dirac point $K$ in the basis $\left\{\phi_{1}^{\uparrow},\phi_{1}^{\downarrow},\phi_{4}^{\uparrow},\phi_{4}^{\downarrow}\right\}$
\begin{eqnarray}\label{HSO1st}
H_{so}^{1st}=\left(\begin{array}{cccc}
-\lambda_{so}^{1st} & 0 & 0 & 0\\
0 & \lambda_{so}^{1st} & 0 & 0\\
0 & 0 & \lambda_{so}^{1st} & 0\\
0 & 0 & 0 & -\lambda_{so}^{1st}
\end{array}\right),
\end{eqnarray}
\begin{equation}\label{lambda1st}
\lambda_{so}^{1st}\equiv\frac{\xi_{0}}{2}| u_{31}|^{2},
\end{equation}
where $u_{31}$ is the corresponding matrix element in $U_A$. In the following, we explain the microscopic mechanism leading to the above equation. The intrinsic effective first order SOC can be summarized as:
\begin{equation}\label{1st}
\begin{split}
&|p_{z\uparrow}^{A}\rangle\overset{V}{\longrightarrow}|p_{-\uparrow}^{B}\rangle\overset{-\frac{\xi_{0}}{2}}{\longrightarrow}|p_{-\uparrow}^{B}\rangle\overset{V}{\longrightarrow}|p_{z\uparrow}^{A}\rangle, \\
&|p_{z\downarrow}^{A}\rangle\overset{V}{\longrightarrow}|p_{-\downarrow}^{B}\rangle\overset{\frac{\xi_{0}}{2}}{\longrightarrow}|p_{-\downarrow}^{B}\rangle\overset{V}{\longrightarrow}|p_{z\downarrow}^{A}\rangle, \\
&|p_{z\uparrow}^{B}\rangle\overset{V}{\longrightarrow}|p_{+\uparrow}^{A}\rangle\overset{\frac{\xi_{0}}{2}}{\longrightarrow}|p_{+\uparrow}^{A}\rangle\overset{V}{\longrightarrow}|p_{z\uparrow}^{B}\rangle,
\\
&|p_{z\downarrow}^{B}\rangle\overset{V}{\longrightarrow}|p_{+\downarrow}^{A}\rangle\overset{-\frac{\xi_{0}}{2}}{\longrightarrow}|p_{+\downarrow}^{A}\rangle\overset{V}{\longrightarrow}|p_{z\downarrow}^{B}\rangle,
\end{split}
\end{equation}
where $V$ means the nearest neighbor direct hopping, $\xi_{0}$ represents the atomic intrinsic spin-orbit interaction strength. The whole process can be divided to three steps. Take $p_z^A$ for example. Firstly, due to the low-buckled structure , $p_z^{A}$ couples to $p_{-}^{B}$(see in Eq.~(\ref{H_{A}}) and Eq.~(\ref{phi_1,4}). Carrier in $p_z^{A}$ orbit directly hops to the nearest neighbor $p_{-}^B$ orbit. Secondly, when the atomic intrinsic SOC is introduced, the energy of $p_{-}^{B}$ will split with spin up carrier shifting $-\frac{\xi_0}{2}$ while spin down carrier shifting $\frac{\xi_0}{2}$. The third, carrier in $p_{-}^{B}$ directly hops to another nearest neighbor $p_z^A$ orbit. The SOC process in the $p_z^B$ is analogous to that of $p_z^A$ except that $p_z^B$ orbit couples to the $p_{+}^{A}$ orbit. The difference leads to the opposite magnitude of effective SOC . During the whole SOC process, the atomic intrinsic SOC takes effect for only one time. Therefore, the effective SOC is proportional to $\xi_0$. A brief sketch of the process is shown in Fig.~\ref{fig:gap}(b).

For most of time, we focus on the low-buckled geometry with small $V_3$. According to the Eqs.~(\ref{U_A})(\ref{eigenvalue})(\ref{lambda1st}), $\lambda_{so}^{1st}$ reads
\begin{equation}\label{lambda_{so}^{1st}}
\begin{split}
& \lambda_{so}^{1st} = \frac{\xi_{0}}{2}\frac{\varepsilon_1 ^2}{\alpha_{1}^{2} V_3^2} \approx \frac{\xi_{0}}{2}\frac{2}{9}\frac{\Delta^{2}\left(V_{pp\pi}-V_{pp\sigma}\right)^{2}}{V_{sp\sigma}^{4}} \\
& \times\frac{\cot^{2}\theta}{1+\frac{\cos^{2}\theta\left(V_{pp\pi}-V_{pp\sigma}\right)^{2}}{V_{sp\sigma}^{2}}\left(1+\frac{2}{9}\frac{\Delta^{2}}{\sin^{2}\theta V_{sp\sigma}^{2}}\right)}.
\end{split}
\end{equation}
Especially, when low-buckled geometry returns to planar structure such as graphene($\theta=90\degree$), the above formula becomes $ \lambda_{so}^{1st}=0 $ and the first order SOC vanishes. Physically, when $\theta=90\degree$, $p_z^A$ orbit is orthogonal with $p_x^{B}$ and $p_y^B$ orbits. Therefore, the directly hoping from $p_z^{A}$ to $p_-^B$ is completely forbidden. The SOC process described in Eq.~\eqref{1st} cannot happen.

The effective second order spin-orbit interaction is also deduced, whose detail derivation is described in Appendix C. Here, we do not intend to repeat them again but just to quote some expressions there Eq.~\eqref{H2st}
\begin{equation}\label{Hso2stgeneral}
H_{so}^{2st}=-H_{n}\left(H_{\sigma}-\varepsilon_{1}\right)^{-1}H_{n}^{\dagger},
\end{equation}
where $H_{n}$ take from the first row to the fourth row and the fifth column to the sixteenth column of the above $H_{so}^{'}$, $H_{\sigma}$ is the direct product matrix between the lower right $6\times6$ diagonal matrix of $H_{0}^{'}$(Eq.~\eqref{H0'}) and $2\times2$ identity matrix, $\varepsilon_{1}$ is the eigenvalue aforementioned. When $V_{3}$ is small, the effective second order SOC Hamiltonian reads at the Dirac point $K$ in the basis $\left\{\phi_{1}^{\uparrow},\phi_{1}^{\downarrow},\phi_{4}^{\uparrow},\phi_{4}^{\downarrow}\right\}$
\begin{eqnarray}\label{Hso2st}
H_{so}^{2st}\simeq-\lambda_{so}^{2st}+\left(\begin{array}{cccc}
-\lambda_{so}^{2st} & 0 & 0 & 0\\
0 & \lambda_{so}^{2st} & 0 & 0\\
0 & 0 & \lambda_{so}^{2st} & 0\\
0 & 0 & 0 & -\lambda_{so}^{2st}
\end{array}\right),
\end{eqnarray}
\begin{equation}\label{lambda_{so}^{2st}}
\begin{split}
 \lambda_{so}^{2st} &\equiv (\frac{\xi_{0}}{2})^{2} [\frac{|{ u_{11}u_{32}}-{ u_{31} u_{12}}|^2}{\varepsilon_2-\varepsilon_1}+\frac{|{ u_{11}u_{33}}-{ u_{31} u_{13}}|^2}{\varepsilon_3-\varepsilon_1}\\
&+ \frac{ u_{11}^2 \varepsilon_1}{\varepsilon_1^2-V_1^2}]  \approx \left(\frac{\xi_{0}}{2}\right)^{2}\frac{2}{9}\frac{-\Delta}{\sin^{2}\theta V_{sp\sigma}^{2}}.
\end{split}
\end{equation}

We analyze the microscopic mechanism for the $\lambda_{so}^{2st}$. Without SOC, the low-energy $H_{\pi}$ and the high energy $H_{\sigma}$ are decoupled. However, in the presence of the atomic intrinsic SOC, $H_{\pi}$ and $H_{\sigma}$ are coupled together. A detail analysis is shown that $\lambda_{so}^{2st}$ can be summarized as the process as:
\begin{equation}
\begin{split}
&|p_{z\uparrow}^{A}\rangle\overset{\xi_{0}/\sqrt{2}}{\longrightarrow}|p_{+\downarrow}^{A}\rangle\overset{V}{\longrightarrow}|s_{\downarrow}^{B}\rangle\overset{V}{\longrightarrow}|p_{+\downarrow}^{A}\rangle\overset{\xi_{0}/\sqrt{2}}{\longrightarrow}|p_{z\uparrow}^{A}\rangle,
\\
&|p_{z\downarrow}^{B}\rangle\overset{\xi_{0}/\sqrt{2}}{\longrightarrow}|p_{-\uparrow}^{B}\rangle\overset{V}{\longrightarrow}|s_{\uparrow}^{A}\rangle\overset{V}{\longrightarrow}|p_{-\uparrow}^{B}\rangle\overset{\xi_{0}/\sqrt{2}}{\longrightarrow}|p_{z\downarrow}^{B}\rangle,
\end{split}
\end{equation}
where $V$ means the nearest neighbor direct hopping, $\xi_{0}$ represents the atomic intrinsic spin-orbit interaction strength. During the process, the atomic intrinsic SOC takes effect for twice. Thus, the effective SOC is the second order of $\xi_0$. A brief sketch of the process is shown in Fig.~\ref{fig:gap}(c). We notice that in graphene, the second order $\lambda_{so}^{2st}$ is the leading order of effective SOC and had been studied in references\cite{huertas-hernando_spin-orbit_2006,yao_spin-orbit_2007,min_intrinsic_2006,castro_neto_electronic_2009}.

\subsection{Intrinsic Rashba SOC in silicene}

The extrinsic Rashba SOC in graphene is due to a perpendicular electronic field or interaction with a substrate which breaks the mirror symmetry, while the intrinsic Rashba SOC in silicene is due to its own low-buckled geometry. Around the $K$ point, the Hamiltonian containing $\vec{k}$ deviation from the $K$ point in the representation $\{p_{z}^{A},p_{z}^{B},p_{y}^{A},p_{x}^{A},s^{A},p_{y}^{B},p_{x}^{B},s^{B}\}$ reads
\begin{equation}
H_{0}\left(k\right)=\delta H_{0}\left(k\right)+H_{0},
\end{equation}
where $H_{0}$ is given in Eq.~\eqref{H0}.
\begin{eqnarray}
\delta H_{0}\left(k\right)=\left(\begin{array}{cc}
\delta h_{11} & \delta h_{12}\\
\delta h_{12}^{\dagger} & \delta h_{22}
\end{array}\right),
\end{eqnarray}
\begin{eqnarray*}
\delta h_{11}=\left(\begin{array}{cccc}
0 & -v_{4}k_{+} & 0 & 0\\
-v_{4}k_{-} & 0 & v_{3}k_{+} & -iv_{3}k_{+}\\
0 & v_{3}k_{-} & 0 & 0\\
0 & iv_{3}k_{-} & 0 & 0
\end{array}\right),
\end{eqnarray*}
\begin{eqnarray*}
\delta h_{22}=\left(\begin{array}{cccc}
0 & -v_{6}k_{-} & -iv_{6}k_{-} & v_{7}k_{+}\\
-v_{6}k_{+} & 0 & 0 & 0\\
iv_{6}k_{+} & 0 & 0 & 0\\
v_{7}k_{-} & 0 & 0 & 0
\end{array}\right),
\end{eqnarray*}
\begin{eqnarray*}
\delta h_{12}=\left(\begin{array}{cccc}
0 & v_{3}k_{-} & iv_{3}k_{-} & v_{5}k_{+}\\
-v_{5}k_{-} & 0 & 0 & 0\\
0 & v_{2}k_{+}-v_{1}k_{-} & iv_{1}k_{-} & v_{6}k_{-}\\
0 & iv_{1}k_{-} & v_{2}k_{+}+v_{1}k_{-} & iv_{6}k_{-}
\end{array}\right),
\end{eqnarray*}
\begin{equation*}
\begin{split}
& v_{1}\equiv\frac{\sqrt{3}}{8}\sin^{2}\theta\left(V_{pp\pi}-V_{pp\sigma}\right)a,    \\
& v_{2}\equiv\frac{\sqrt{3}}{4}\left[\sin^{2}\theta\left(V_{pp\pi}-V_{pp\sigma}\right)-2V_{pp\pi}\right]a,    \\
& v_{3}\equiv\frac{\sqrt{3}}{4}\sin\theta\cos\theta\left(V_{pp\pi}-V_{pp\sigma}\right)a , \\
& v_{4}\equiv\frac{\sqrt{3}}{2}\left(V_{pp\pi}\sin^{2}\theta+V_{pp\sigma}\cos^{2}\theta\right)a, \\
& v_{5}\equiv\frac{\sqrt{3}}{2}\cos\theta V_{sp\sigma}a, v_{6}\equiv\frac{\sqrt{3}}{4}\sin\theta V_{sp\sigma}a, v_{7}\equiv-\frac{\sqrt{3}}{2}V_{ss\sigma}a.
\end{split}
\end{equation*}
Through the unitary transformation matrix $U$ (Eq.~\eqref{U}), in the representation $\left\{ \phi_{1},\phi_{4},\phi_{2},\phi_{5},\phi_{3},\phi_{6},\phi_{7},\phi_{8}\right\}$, we have
\begin{equation}
H_{0}'\left(k\right) \longrightarrow H_{0}'\left(k\right)=U^{\dagger}H_{0}\left(k\right)U= \delta H_{0}^{'}\left(k\right)+H_{0}^{'},
\end{equation}
where $H_{0}^{'}$ is given in Eq.~\eqref{H0'}. We mainly focus on the terms containing $\vec{k}$ deviation from the $K$ point
\begin{equation}
H^{'}\left(k\right) \longrightarrow H^{'}\left(k\right)\equiv H_{0}^{'}\left(k\right)\otimes I_{2}+H_{so}^{'},
\end{equation}
where $H_{so}^{'}$ is given in Eq.~\eqref{Hso'matrix}. According to the Eq.~\eqref{H2st}, the total second order Hamiltonian reads
\begin{equation}
H_{eff}^{'}(k)=-H_{non}\left(k\right)\left(H_{\sigma}-\varepsilon_{1}\right)^{-1}H_{non}^{\dagger}\left(k\right),
\end{equation}
where $H_{non}\left(k\right)$ takes from the first row to the fourth row and the fifth column to the sixteenth column of $H^{'}\left(k\right)$. The Hamiltonian $H_{eff}^{'}(k)$ can be divide into two parts
\begin{equation}
H_{eff}^{'}(k)=H_{so}^{2st}+H_{R}\left(k\right),
\end{equation}
where $H_{so}^{2st}$ is given in Eq.~\eqref{Hso2st}. $H_{R}\left(k\right)$ is intrinsic Rashba SOC in silicene, which can be written as around the Dirac point $K$ in the basis $\left\{ \phi_{1}^{\uparrow},\phi_{1}^{\downarrow},\phi_{4}^{\uparrow},\phi_{4}^{\downarrow}\right\}$
\begin{eqnarray}\label{H_{R}}
H_{R}\left(k\right)=\left(\begin{array}{cccc}
0 & -i\lambda_{R}ak_{-} & 0 & 0\\
i\lambda_{R}ak_{+} & 0 & 0 & 0\\
0 & 0 & 0 & i\lambda_{R}ak_{-}\\
0 & 0 & -i\lambda_{R}ak_{+} & 0
\end{array}\right),
\nonumber\\
&&
\end{eqnarray}
where the purely real $\lambda_{R}$ reads
\begin{equation}\label{lambda_{R}}
\begin{split}
& \lambda_{R}=\frac{i\xi_{0}}{\sqrt{2}}\frac{u_{11}u_{32}-u_{31}u_{12}}{\left(\varepsilon_{2}-\varepsilon_{1}\right)a}\times\\
& \left[\left(u_{12}u_{21}+u_{22}u_{11}\right)v_{5}+u_{22}u_{21}v_{7}+u_{12}u_{11}v_{4}-2u_{32}u_{31}v_{1}\right] \\
& +\frac{i\xi_{0}}{\sqrt{2}}\frac{u_{11}u_{33}-u_{31}u_{13}}{\left(\varepsilon_{3}-\varepsilon_{1}\right)a}\times \\
& \left[\left(u_{13}u_{21}+u_{23}u_{11}\right)v_{5}+u_{23}u_{21}v_{7}+u_{13}u_{11}v_{4}-2u_{33}u_{31}v_{1}\right]\\
& +\xi_{0}\frac{u_{11}\left(u_{11}v_{3}-u_{21}v_{6}-\frac{i}{\sqrt{2}}u_{31}v_{2}\right)}{2\left(V_{1}+\varepsilon_{1}\right)a}\\
& -\xi_{0}\frac{u_{11}\left(-u_{11}v_{3}+u_{21}v_{6}-\frac{i}{\sqrt{2}}u_{31}v_{2}\right)}{2\left(V_{1}-\varepsilon_{1}\right)a}.
\end{split}
\end{equation}
From the above equations, we know that $H_{R}\left(k\right)$ is exactly zero at Dirac point $K$, while $H_{R}\left(k\right)$ has nonzero value with $\vec{k}$ deviation from the $K$ point. Moreover, when the structure returns to the planar structure, $\theta =90\degree, \lambda_R=0$, the intrinsic Rashba SOC vanishes even when $\vec{k}$ deviating from $K$. Therefore the intrinsic Rashba is entirely caused by the low-buckled geometry. The intrinsic Rashba SOC is quite different from the extrinsic Rashba SOC, which arising from a perpendicular electronic field or interaction with a substrate leading to mirror symmetry broken in some direction, has finite magnitude at Dirac point K.

\section{Result and DISCUSSION}

Finally, in combination with Eqs.~\eqref{HK}\eqref{HSO1st}\eqref{Hso2st}\eqref{H_{R}}, we obtain the entire low-energy effective Hamiltonian around Dirac $K$ acting on the low-energy states $\phi_{1}$ and $\phi_{4}$
\begin{equation}\label{HeffTB}
\begin{split}
& H_{K}^{eff} \left(\theta\right)=H_{K}\otimes I_{2}+H_{so}^{1st}+H_{so}^{2st}+H_{R}\left(k\right) \\
& =(\varepsilon_{1}-\lambda_{so}^{2st})I_4+\left(\begin{array}{cc}
h_{11} & v_{F}k_{+}\\
v_{F}k_{-} & -h_{11}
\end{array}\right),
\end{split}
\end{equation}
\begin{equation*}
h_{11}\equiv-\lambda_{so}\sigma_{z}-a\lambda_{R}\left(k_{y}\sigma_{x}-k_{x}\sigma_{y}\right),
\end{equation*}
where $I_{2}$ is $2 \times 2$ identity matrix for the spin degree of freedom, $I_{4}$ is $4 \times 4$ identity matrix and $\lambda_{so}=\lambda_{so}^{1st}+\lambda_{so}^{2st}$, $v_{F}$ is given in Eq.~\eqref{vf}. Through the time-reversal operation, the entire Low-energy effective Hamiltonian around Dirac $K^*$ reads
\begin{equation}
H_{K^*}^{eff} \left(\theta\right)=(\varepsilon_{1}-\lambda_{so}^{2st}) I_4  +\left(\begin{array}{cc}
-h_{11} & v_{F}k_{-}\\
v_{F}k_{+} & h_{11}
\end{array}\right).
\end{equation}
The effective Hamiltonian deduced from atomic tight-binding method has the similar formulas as that from the symmetry aspects. Comparing Eq.~\eqref{HeffTB} and Eq.~\eqref{HeffSmy}, we obtain
\begin{equation}\label{model parameters}
t=\frac{2\sqrt{3}v_{F}}{3a}, t_{2}=\frac{\lambda_{so}}{3\sqrt{3}}, t_{1}=\frac{2}{3}\lambda_{R} .
\end{equation}
The above parameters $t, t_{2}, t_{1}$ are undetermined in the second nearest neighbor tight-binding model Eqs.~(\ref{HSmy}) and (\ref{HeffSmy}) from the above symmetry analysis. Here, in combination with Eqs.~\eqref{vf}\eqref{lambda_{so}^{1st}}\eqref{lambda_{so}^{2st}}\eqref{lambda_{R}}\eqref{model parameters}, we can not only give their explicit expressions, but also specify the magnitudes of the three parameters through $v_{F}$, $\lambda_{so}$($\lambda_{so}=\lambda_{so}^{1st}+\lambda_{so}^{2st}$), and $\lambda_{R}$, whose values are presented in Table ~\ref{tab:application}.

In the following, we discuss the physic meanings of our obtained low-energy effective Hamiltonian. First of all, the low-energy effective Hamiltonian is analogous to the first QSHE proposal in graphene except the intrinsic Rashba SOC term $H_{R}\left(k\right)$~\cite{kane_z2_2005}. The SOC inducing mass term to Hamiltonian opens gap at the Dirac points. Moreover, from K to $K^{*}$ the mass term changes its sign and the band is inverted. Therefore, the low-buckled silicene is also the QSHE system. The QSHE can be observed experimentally when the Fermi energy locates inside the gap and the temperature is below the minimal energy gap. The existence of the QSHE in silicene has been studied in our recent work using the first-principles method combined with the direct $Z_2$ calculation~\cite{liu_quantum_2011}.

\begin{figure}
\includegraphics[width=3.5in]{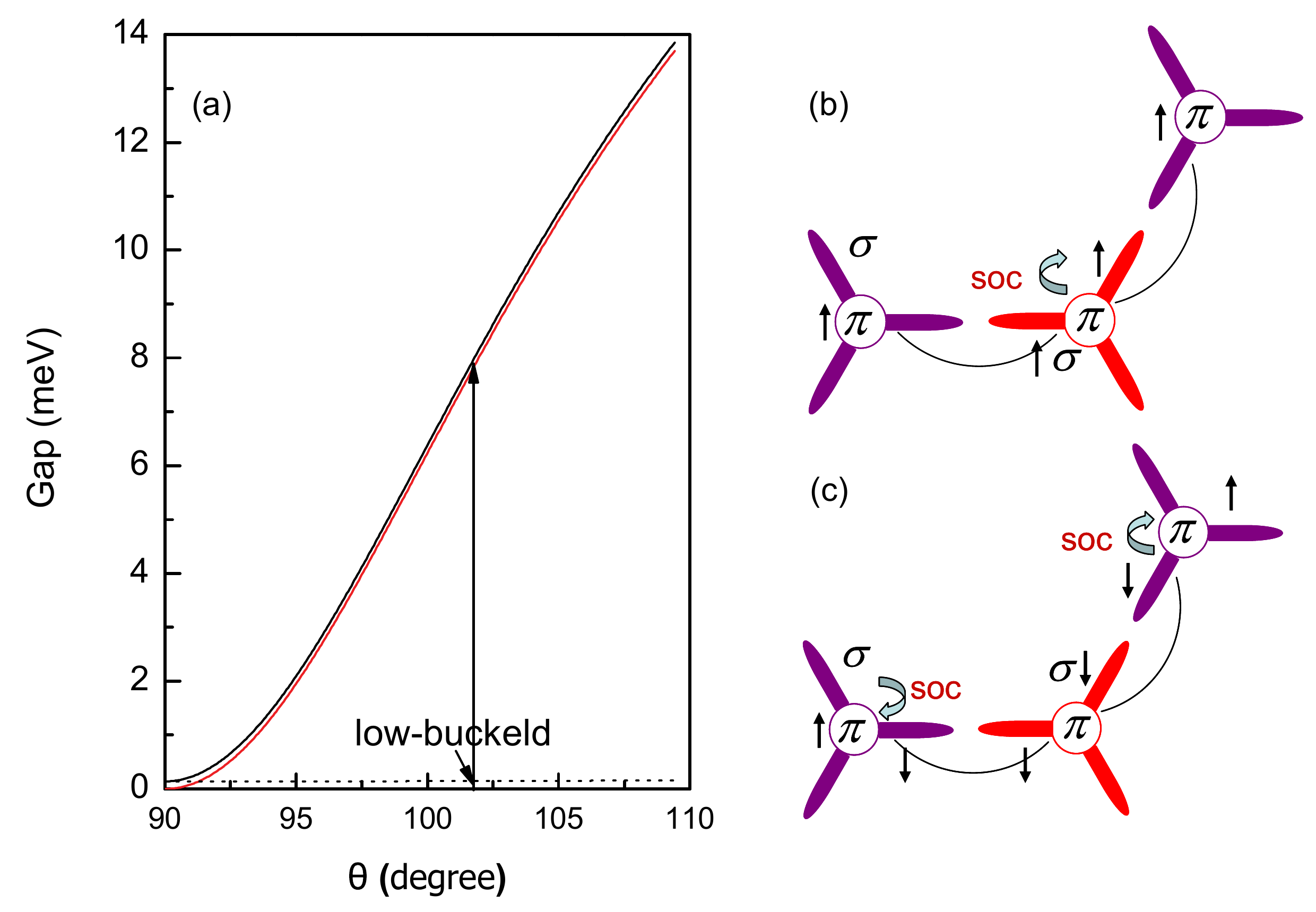}
\caption{(color online). (a) The variation of gap opened by SOC at Dirac point with the angle $\theta$ for silicene. Black line marks the total gap. Red or gray line and dot line mean gaps opened by the first order and second order SOC respectively. (b), (c) Brief sketches of the intrinsic effective first and second order spin orbit interaction.}\label{fig:gap}
\end{figure}

Secondly, the energy gap in low-buckled  silicene is much larger than that in graphene. The Eq.~\eqref{HeffTB} results in a spectrum $E(\vec{k})=\pm\sqrt{\left(v_{F}^{2}+a^{2}\lambda_{R}^{2}\right)k^{2}+\lambda_{so}^{2}} $. Therefore, the energy gap is $2\lambda_{so}$ at the Dirac points. Due to the low-buckled geometry, not only the second order SOC $\lambda_{so}^{2st}$ but also the much larger first order SOC $\lambda_{so}^{1st}$ exist. In Fig.~\ref{fig:gap}, we show the variation of gap with the angle $\theta$. With $\theta$ deviating from $90\degree$, the gap induced by $\lambda_{so}^{2st}$ for silicene is nearly unchanged while the gap induced by $\lambda_{so}^{1st}$ increases rapidly. The larger is the angle, the greater is the gap. Especially, the gap can reach to several meV for just little buckled, therefore the QSHE can be observed in an experimental observable temperature regime.

Thirdly, due to the low-buckled geometry, the effective Hamiltonian also contains the intrinsic Rashba SOC term. Such term leads to interesting properties. On the one hand, since it vanishes at Dirac point, the minimal bulk energy gap $2\lambda_{so}$ will not be affected by the intrinsic Rashba SOC. Therefore, it does not diminish the temperature window for experimentally observing the QSHE in silicene. On the other hand, due to the nonzero values of the Rashba SOC term, spin is not a good quantum number. Thus, the spin Hall conductance is no longer quantized in silicene. The intrinsic Rashba SOC is entirely different from the extrinsic Rashba SOC, which has finite value at Dirac point K, can destroy the QSHE~\cite{kane_z2_2005}. The presence of intrinsic Rashba SOC may provide a way to manipulate the spin in silicene without destroying its QSHE state.

Fourthly, the entire low-energy effective Hamiltonian applies to not only the silicene itself but also the low-buckled counterparts of graphene for other group IVA elements Ge and Sn, as well as graphene with the planar geometry. These different structures correspond to the different angles $\theta$. Therefore, in this sense, the effective Hamiltonian is quite general.
\begin{table*}
\caption{An application of the low-energy effective Hamiltonian (Eq.~\eqref{HeffTB}). The terms of this general low-energy effective Hamiltonian for these different systems corresponding to the different angles $\theta$ of the lowest energy geometry are given. The lattice constants $a$ with unit \AA \ and angles $\theta$ for the lowest energy geometry are obtained from the first-principles calculation. $\lambda_{so}^{1st}$, $\lambda_{so}^{2st}$, and $\lambda_{R}$ caused by low-buckled geometry and SOC, with units $meV$ at the Dirac point $K$ are presented obtained from our tight-binding model using hopping parameters in Table~\ref{tab:hopping parameters}. The gap with unit $meV$ opened by SOC at Dirac point $K$ from first-principles and the current tight-binding method is presented. We also give the carrier Fermi velocity $v_{F}$ with unit $10^5m/s$ around Dirac point $K$ from first-principles and the current tight-binding method.}\label{tab:application}
\begin{ruledtabular}
\begin{tabular}{cccccccccc}
\multicolumn{1}{c}{\textrm{System}} &
\multicolumn{1}{c}{\textrm{$a$(\AA)}} &
\multicolumn{1}{c}{\textrm{$\theta$}} &
\multicolumn{1}{c}{\textrm{$\lambda_{so}^{1st} \left(meV \right)$}} &
\multicolumn{1}{c}{\textrm{$ \lambda_{so}^{2st}$}} &
\multicolumn{1}{c}{\textrm{$\lambda_{R}$}}&
\multicolumn{1}{c}{\textrm{$Gap \left(meV \right)(TB)$}} &
\multicolumn{1}{c}{\textrm{$Gap(FP)$}} &
\multicolumn{1}{c}{\textrm{$v_{F} \left(10^5m/s \right)(TB) $}} &
\multicolumn{1}{c}{\textrm{$v_{F}(FP)$}} \\
\hline
graphene   &2.46 & $90\degree$       & 0    & 1.3$\times10^{-3}$   & 0    &2.6$\times$$10^{-3}$  &0.8$\times$$10^{-3}$\footnotemark[1]  & 9.80 & 8.46 \\
silicene   &3.86 & $101.7\degree$    & 3.9  & 7.3$\times10^{-2}$   & 0.7  &7.9                   &1.55\footnotemark[2]                  & 5.52 & 5.42 \\
ge(licene) &4.02 & $106.5\degree$    & 43   & 3.3                  & 10.7 &93                    &23.9\footnotemark[2]                  & 4.57 & 5.24 \\
sn(licene) &4.70 & $107.1\degree$    & 29.9 & 34.5                 & 9.5  &129                   &73.5                                  & 4.85 & 4.70 \\
\end{tabular}
\end{ruledtabular}
\footnotetext[1]{Reference~\onlinecite{yao_spin-orbit_2007}.}
\footnotetext[2]{Reference~\onlinecite{liu_quantum_2011}.}
\end{table*}

In Table~\ref{tab:application}, $\lambda_{so}^{1st}$, $\lambda_{so}^{2st}$, $\lambda_{R}$, the gap, and $v_{F}$ caused by SOC at Dirac point $K$ in graphene, silicene, gelicene and snlicene, similarly corresponding to two-dimensional low-buckled Ge and Sn, are obtained from tight-binding method by using the typical parameters values from Table ~\ref{tab:hopping parameters}. Notice that $\lambda_{so}^{2st}$ is slightly larger than $\lambda_{so}^{1st}$ due to the huge SOC strength with the magnitude of $eV$ in snlicene, while $\lambda_{so}^{2st}$ is much smaller than $\lambda_{so}^{1st}$ in other systems. For comparison, from the first-principles method, we present the corresponding gaps too, which agree with our tight-binding method results in order of magnitude. We also give the carrier Fermi velocity $v_{F}$ around Dirac point $K$ from first-principles and the current tight-binding method. Since we only focus on the low-buckled geometry, our calculation shows that the carrier Fermi velocity does not significantly change with $\theta$.

Notice that those bond parameters presented in Table~\ref{tab:hopping parameters} used in Table~\ref{tab:application} come from the corresponding diamond structure($sp3$ hybridization) actually except the graphene($sp2$) case. However, considering that low-bulked structures are more closely to $sp2$ hybridization and bond parameters of $sp2$ hybridization will be a little different from those of $sp3$ hybridization, thus, through slight improvement of these bond parameters we expect that the tight-binding gap can better match with the first-principles results.

\section{Summary}

In summary, based on the symmetry aspects and tight-binding method combined with first-principles calculation, we derived the low-energy effective Hamiltonian for silicene, which is very general because this Hamiltonian applies to not only the silicene itself but also the low-buckled counterparts of graphene for other group IVA element Ge and Sn, as well as graphene when the structure returns to the planar geometry. The low-energy effective Hamiltonian is indeed QSHE with its form similar to Kane-Mele's first graphene QSHE Hamiltonian except the intrinsic Rashba SOC term $H_{R}\left(k\right)$. However, the effective SOC in low-buckled geometry is actually first order in the atomic intrinsic SOC strength $\xi_{0}$, while the planar structure in graphene leads to the vanishing of the leading-order contribution. Therefore, silicene as well as low-buckled counterparts of graphene for other group IVA elements Ge and Sn has much larger gap opened by effective SOC at Dirac point than graphene due to low-buckled geometry and larger atomic intrinsic SOC strength. Further, the larger is the angle, the greater is the gap. Therefore, QSHE can be observed in an experimentally accessible low temperature regime in these low-buckled systems.  In addition, the Rashba SOC in silicene is intrinsic due to its own low-buckled geometry, which vanishes at Dirac point $K$, while has nonzero value with $\vec{k}$ deviation from the $K$ point.  As a result, though spin Hall conductance is not quantized, the QSHE in the silicene is robust against to such intrinsic Rashba SOC. This is entirely different from the extrinsic Rashba SOC due to a perpendicular electronic field or interaction with a substrate, which is independent of $\vec{k}$, has finite value at Dirac points, and can destroy the QSHE.

\begin{acknowledgments}
We would like to thank Qian Niu, Junren Shi and Haiwen Liu for helpful discussions. This work was supported by NSF of China (Grant No. 10974231) , the MOST Project of China (Grants No.2007CB925000 and 2011CBA00100), CPSF No. 20100480147, and 985 Program of Peking University.
\end{acknowledgments}

\appendix

\section{$H_0$ Matrix}
In the representation $\left\{p_{z}^{A},p_{z}^{B},p_{y}^{A},p_{x}^{A},s^{A},p_{y}^{B},p_{x}^{B},s^{B}\right\}$ the total Hamiltonian reads
\begin{eqnarray*}
  H_{0}=\left(
  \begin{array}{cc}
    \mathbf{H}_{\pi} & \mathbf{H_{n}}\\
    \mathbf{H_{n}}^{\dagger} & \mathbf{H}_{\sigma}
  \end{array}
  \right),\label{originalH}
\end{eqnarray*}
\begin{eqnarray}
  \mathbf{H}_{\sigma}=\left(
  \begin{array}{cc}
    \mathbf{E} & \mathbf{T}\\
    \mathbf{T}^{\dagger} & \mathbf{E}
  \end{array}
  \right).
\end{eqnarray}
Here, $\mathbf{H}_{\pi}$ and $\mathbf{H}_{\sigma}$ are $2\times2$ and $6\times6$ matrices, respectively. The non-diagonal block $\mathbf{H_{n}}$ coupling $\mathbf{H}_{\pi}$ and $\mathbf{H}_{\sigma}$ is $2\times6$ matrix. In the following derivation, the energy level of $3p$ orbital is set as energy zero point. The matrix $\mathbf{E}$ describes the on-site energy of different atomic orbitals, which can be written as
\begin{eqnarray}
  \mathbf{E}=\left(\begin{array}{ccc}
0 & 0 & 0\\
0 & 0 & 0\\
0 & 0 & \Delta
\end{array}\right),
\end{eqnarray}
where $\Delta$ is the energy difference between the $3s$ and $3p$ orbitals. Actually, here we have assumed these basis are orthogonal when centered on different sites for the sake of simplicity. We choose the coordinate system in which the unit cell has primitive vectors
\begin{equation}
\vec{a}_{1}=a\left(\frac{1}{2},\frac{\sqrt{3}}{2}\right), \vec{a}_{2}=a\left(-\frac{1}{2},\frac{\sqrt{3}}{2}\right).
\end{equation}
The lattice constant $a$ is defined as the nearest distance of lattice point at the same sublattice, which is $3.86$ \AA \  for silicene from our first principles calculation~\cite{liu_quantum_2011}. The three nearest neighbor translation vectors are
\begin{equation}
\begin{split}
& \vec{d_{1}}=\frac{a}{\sqrt{3}}\left(\frac{\sqrt{3}}{2},\frac{1}{2},\cot\theta\right), \\
& \vec{d_{2}}=\frac{a}{\sqrt{3}}\left(-\frac{\sqrt{3}}{2},\frac{1}{2},\cot\theta\right) , \\
& \vec{d_{3}}=\frac{a}{\sqrt{3}}\left(0,-1,\cot\theta\right).
\end{split}
\end{equation}
As shown in Fig.~\ref{fig:geometry}, the angle $\theta$ is defined as being between the Si-Si bond and the $z$ direction normal to the plane. The corresponding reciprocal lattice vectors are
\begin{equation}
\vec{b}_{1}=\frac{2\pi}{a}\left(1,\frac{\sqrt{3}}{3}\right) , \vec{b}_{2}=\frac{2\pi}{a}\left(-1,\frac{\sqrt{3}}{3}\right).
\end{equation}
The Dirac point $K$ is chosen to be $\vec{K}=\frac{1}{3}\left(\vec{b}_{1}-\vec{b}_{2}\right)=\left(\frac{4\pi}{3a},0\right)$, as well as $K^*$=$-K$. The matrix $\mathbf{T}$ describes the hopping between two sublattices, which is given in Table~\ref{tab:Slater} by the Slater-Koster formula\cite{slater_simplified_1954}. In Table~\ref{tab:Slater}, the four bond parameters $V_{ss\sigma}$, $V_{sp\sigma}$, $V_{pp\sigma}$ and $V_{pp\pi}$ correspond to the $\sigma$ and $\pi$ bond formed by $3s$ and $3p$ orbitals, whose numerical values given in Table~\ref{tab:hopping parameters} specify our model quantitatively. The hopping matrix elements in the momentum space read
\begin{equation} \label{hopping matrix elements}
t\left(k\right)=\sum_{i=1}^{3}t(\vec{d_{i}})e^{i\vec{k}\cdot\vec{d_{i}}}.
\end{equation}
Therefore, the matrix $\mathbf{T}$ and $\mathbf{H_{n}}$ at the Dirac point $K$ can be written as
\begin{eqnarray}
\mathbf{T}=\left(\begin{array}{ccc}
-V_{1}^{'} & -iV_{1}^{'} & V_{2}^{'}\\
-iV_{1}^{'} & V_{1}^{'} & -iV_{2}^{'}\\
-V_{2}^{'} & iV_{2}^{'} & 0
\end{array}\right),
\end{eqnarray}
\begin{eqnarray}
\mathbf{H_{n}}=\left(\begin{array}{cccccc}
0 & 0 & 0 & V_{3}^{'} & -iV_{3}^{'} & 0\\
V_{3}^{'} & iV_{3}^{'} & 0 & 0 & 0 & 0
\end{array}\right),
\end{eqnarray}
\begin{equation*}
\begin{split}
& V_{1}^{'}\equiv\frac{3}{4}\sin^{2}\theta\left(V_{pp\pi}-V_{pp\sigma}\right), \\
& V_{2}^{'}\equiv\frac{3}{2}\sin\theta V_{sp\sigma}, \\
& V_{3}^{'}\equiv\frac{3}{2}\sin\theta\cos\theta\left(V_{pp\pi}-V_{pp\sigma}\right).
\end{split}
\end{equation*}
The matrix $\mathbf{H}_{\pi}$ at the Dirac point $K$ reads
\begin{eqnarray}
\mathbf{H}_{\pi}=\left(\begin{array}{cc}
0 & 0\\
0 & 0
\end{array}\right).
\end{eqnarray}
Consequently, Hamiltonian $H_0$ is obtained.
\begin{table}
\caption{The matrix elements for the nearest neighbor hopping between $s$ and $p$ orbitals are considered as functions of the direction cosine $l$, $m$ and $n$ of the vector from the left orbital to the right orbital. Other matrix elements are found by permuting indices.}\label{tab:Slater}
\begin{ruledtabular}
\begin{tabular}{ccccccc}
& $t_{s,s}$ &  $V_{ss\sigma}$  & $t_{x,x}$ &  $l^2$$V_{pp\sigma}$+(l-$l^2$)$V_{pp\pi}$ \\
& $t_{s,x}$ &  $l$$V_{sp\sigma}$ & $t_{x,y}$ &  $lm$($V_{pp\sigma}$-$V_{pp\pi}$)\\
& $t_{x,s}$ &  -$l$$V_{sp\sigma}$ & $t_{y,z}$ &  $mn$($V_{pp\sigma}$-$V_{pp\pi}$)\\
\end{tabular}
\end{ruledtabular}
\end{table}

\begin{table}
\caption{The magnitudes of hopping parameters. The energy units are $eV$. The strength of SOC $\xi_{0}$ is obtained from first-principles calculation except Sn.}\label{tab:hopping parameters}
\begin{ruledtabular}
\begin{tabular}{ccccccc}
\multicolumn{1}{c}{\textrm{System}} &
\multicolumn{1}{c}{\textrm{$V_{ss\sigma}$}} &
\multicolumn{1}{c}{\textrm{$V_{sp\sigma}$}} &
\multicolumn{1}{c}{\textrm{$V_{pp\sigma}$}} &
\multicolumn{1}{c}{\textrm{$V_{pp\pi}$}} &
\multicolumn{1}{c}{\textrm{$\Delta$}} &
\multicolumn{1}{c}{\textrm{$\xi_{0}$}} \\
\hline
Graphene & -6.769 & 5.580 & 5.037 & -3.033 & -8.868\footnotemark[1] &9$\times$$10^{-3}$\footnotemark[3]  \\
Silicene & -1.93 & 2.54 & 4.47 & -1.12 & -7.03\footnotemark[2] &34$\times$$10^{-3}$\footnotemark[4]\\
Ge(licene) & -1.79 & 2.36 & 4.15 & -1.04 & -8.02\footnotemark[2] &0.196\\
Sn(licene) & -2.6245 & 2.6504 & 1.4926 & -0.7877 & -6.2335\footnotemark[5] &0.8\footnotemark[6]\\
\end{tabular}
\end{ruledtabular}
\footnotetext[1]{Reference~\onlinecite{saito_electronic_1992}.}
\footnotetext[2]{Reference~\onlinecite{harrison_electronic_1989}.}
\footnotetext[3]{Reference~\onlinecite{yao_spin-orbit_2007}.}
\footnotetext[4]{Reference~\onlinecite{liu_quantum_2011}.}
\footnotetext[5]{Reference~\onlinecite{pedersen_tight-binding_2010}.}
\footnotetext[6]{Reference~\onlinecite{PhysRevB.16.790}.}
\end{table}

\section{$H_{so}$ Matrix}
When in center field, Eq.~\eqref{Hsogeneral} reads
\begin{equation}\label{Hso}
H_{so}=\xi_{0}\vec{L}\cdot\vec{s}.
\end{equation}
The above equation can also be written as
\begin{equation}\label{Hsocenter}
H_{so}=\xi_{0}\left(\frac{L_{+}s_{-}+L_{-}s_{+}}{2}+L_{z}s_{z}\right),
\end{equation}
where $s_{\pm }=s_{x}\pm is_{y}$ denote the plus(minus) operator for spin and
$L_{\pm}=L_{x} \pm i L_y$ denote the plus(minus) operator for the angular momentum
in the selected basis.
The SOC on the same atom is taken into account. The concrete SOC term can be obtained by calculating the mean value of the Eq.~\eqref{Hsocenter}. For example, the SOC term between $|p_z\uparrow\rangle $ and $|p_x \downarrow \rangle$ reads $\langle p_z \uparrow |H_{so}|p_x \downarrow \rangle =-\frac{\xi_{0}}{2}$ etc~\cite{liu_model_2010}. During the derivation we may take advantage of the following expressions
\begin{equation}
\begin{split}
& L_{+}|l,m\rangle=\left[l\left(l+1\right)-m\left(m+1\right)\right]^{1/2}|l,m+1\rangle,\\
& L_{-}|l,m\rangle=\left[l\left(l+1\right)-m\left(m-1\right)\right]^{1/2}|l,m-1\rangle,\\
& L_{z}|l,m\rangle=m|l,m\rangle,
\end{split}
\end{equation}
where $l,m$ represent the azimuthal quantum number and magnetic quantum number, respectively. A straightforward calculation leads to the on-site SOC in the representation $\left\{ p_{z}^{A},p_{z}^{B},p_{y}^{A},p_{x}^{A},s^{A},p_{y}^{B},p_{x}^{B},s^{B}\right\}\otimes \{ \uparrow, \downarrow \} $
\begin{eqnarray}
H_{so}=\frac{\xi_{0}}{2}\times h_{so}.
\end{eqnarray}
All elements in $h_{so}$ can be found in the Table~\ref{tab:Hso hopping parameters}.
\begin{table}
  \centering
  \caption{The values of SOC among atomic orbitals that used in $h_{so}$. $A,B$ denote the two distinct sites. The nonzero SOC terms only exist in the same site.  $\sigma_{x,y,z}$ are Pauli matrixes acting on the spin space. O denotes the zero matrix.  }\label{tab:Hso hopping parameters}
\begin{tabular}{ccccc}
  \hline
  \hline
  \ \  \  \ \ \ \ \ \ \ &  \ \  \ $   p_z^{A/B} $ \ \ \ \ \ & \ \ \ $p_y^{A/B} $ \ \ \ \ \ & \ \ \ $p_x^{A/B} $ \ \ \ \ \ & $ \ \ \ s^{A/B}$ \ \ \ \ \ \\ \hline
  $p_z^{A/B}$ & O & $i \sigma_x$ & $-i \sigma_y$ & O \\
  $p_y^{A/B}$ & -$i \sigma_x$ & O & i$\sigma_z$ & O \\
  $p_x^{A/B}$ & $i \sigma_y$ & -$i \sigma_z$ & O & O \\
  $s^{A/B}$ & O & O & O & O \\
  \hline
  \hline
\end{tabular}
\end{table}

\section{the Second order effective Hamiltonian}
In general, the Hamiltonian reads
\begin{eqnarray}
\mathbb{H}=\left(\begin{array}{cc}
H_{\pi} & H_{n}\\
H_{n}^{\dagger} & H_{\sigma}
\end{array}\right).
\end{eqnarray}
We focus on the case: (i) the eigenvalues of $H_{\pi}$ is around energy $\epsilon$ while the eigenvalues of $H_{\sigma}$ is far away from $\epsilon$ (ii) the energy scale of
the non-diagonal block $H_{n}$ is much smaller than the eigenvalue value difference between $H_{\pi}$ and $H_{\sigma}$. The effective Hamiltonian around energy $\epsilon$ (or second order effective Hamiltonian for $H_{\pi}$) can be obtained by the following method~\cite{winkler_spin-orbit_2003}.
$\mathbb{H}$ can be rewritten as
 \begin{eqnarray}
\mathbb{H}\equiv \epsilon I + \mathbb{H}_{0}+\mathbb{H}_{non},
\end{eqnarray}
\begin{eqnarray*}
\mathbb{H}_{0}=\left(\begin{array}{cc}
H_{\pi}- \epsilon  & 0\\
0 & H_{\sigma} - \epsilon
\end{array}\right),  \hspace{1em}   \mathbb{H}_{non}=\left(\begin{array}{cc}
0 & H_{n}\\
H_{n}^{\dagger} & 0
\end{array}\right).
\end{eqnarray*}
For simplicity, we omit the unitary matrix $I$ in the above and the following derivation. In order to obtain the effective Hamiltonian, one may perform a canonical transformation:
\begin{eqnarray*}
\mathbb{H} \longrightarrow H_{S}=e^{-S}\mathbb{H}e^{S},
\end{eqnarray*}
\begin{eqnarray}
S=\left(\begin{array}{cc}
0 & M\\
-M^{\dagger} & 0
\end{array}\right),
\end{eqnarray}
where the matrix $M$ is determined by
\begin{eqnarray}
[\mathbb{H}_{0},S]+\mathbb{H}_{non}=0
\end{eqnarray}
Through simple algebraic derivation, we have
\begin{equation}
(H_{\pi}-\epsilon)M-M(H_{\sigma}-\epsilon)+H_{n}=0.
\end{equation}
Therefore, we can find a recursive expression for $M$
\begin{equation}
\begin{split}
& M=[H_{n}+(H_{\pi}-\epsilon ) M ](H_{\sigma}-\epsilon)^{-1}  \\
& =H_{n}(H_{\sigma}-\epsilon)^{-1}+(H_{\pi}-\epsilon)H_{n}(H_{\sigma}-\epsilon)^{-2}+\cdots.
\end{split}
\end{equation}
We know that in silicene the eigenvalues of $H_{\sigma}-\epsilon$ determined by the energy of $H_{\sigma}$ separated from those of $H_{\pi}$ are of order $eV$ near the $K$ point, while the energy scale of $H_{\pi}-\epsilon$ is nearly zero as well as $H_{n}$ is of order meV for SOC. Therefore the above recursive expression can be written as:

\begin{equation}
M\approx H_{n}(H_{\sigma}-\epsilon)^{-1}.
\end{equation}
The transformed  Hamiltonian has the following approximate form
\begin{eqnarray}
 H_{S}&=& e^{-S}\mathbb{H}e^{S}=\mathbb{H}+[\mathbb{H},S]+\frac{1}{2!}[[\mathbb{H},S],S]+\cdots   \nonumber\\
  &=& \epsilon I+ \mathbb{H}_{0}+\frac{1}{2}[\mathbb{H}_{non},S]+\cdots.
\end{eqnarray}
Up to the second order, the final effective Hamiltonian for $H_{\pi}$ can be written as
\begin{eqnarray}\label{H2st}
H_{eff}& \simeq & H_{\pi}-\frac{1}{2}\left(H_{n}M^{\dagger}+MH_{n}^{\dagger}\right) \nonumber\\
& \simeq & H_{\pi}-H_{n}(H_{\sigma}-\epsilon)^{-1}H_{n}^{\dagger}.
\end{eqnarray}

\bibliographystyle{apsrev}
\bibliography{reference}

\begin{thebibliography}{29}
\expandafter\ifx\csname natexlab\endcsname\relax\def\natexlab#1{#1}\fi
\expandafter\ifx\csname bibnamefont\endcsname\relax
  \def\bibnamefont#1{#1}\fi
\expandafter\ifx\csname bibfnamefont\endcsname\relax
  \def\bibfnamefont#1{#1}\fi
\expandafter\ifx\csname citenamefont\endcsname\relax
  \def\citenamefont#1{#1}\fi
\expandafter\ifx\csname url\endcsname\relax
  \def\url#1{\texttt{#1}}\fi
\expandafter\ifx\csname urlprefix\endcsname\relax\def\urlprefix{URL }\fi
\providecommand{\bibinfo}[2]{#2}
\providecommand{\eprint}[2][]{\url{#2}}

\bibitem[{\citenamefont{Lalmi et~al.}(2010)\citenamefont{Lalmi, Oughaddou,
  Enriquez, Kara, Vizzini, Ealet, and Aufray}}]{lalmi_epitaxial_2010}
\bibinfo{author}{\bibfnamefont{B.}~\bibnamefont{Lalmi}},
  \bibinfo{author}{\bibfnamefont{H.}~\bibnamefont{Oughaddou}},
  \bibinfo{author}{\bibfnamefont{H.}~\bibnamefont{Enriquez}},
  \bibinfo{author}{\bibfnamefont{A.}~\bibnamefont{Kara}},
  \bibinfo{author}{\bibfnamefont{S.}~\bibnamefont{Vizzini}},
  \bibinfo{author}{\bibfnamefont{B.}~\bibnamefont{Ealet}}, \bibnamefont{and}
  \bibinfo{author}{\bibfnamefont{B.}~\bibnamefont{Aufray}},
  \bibinfo{journal}{Appl. Phys. Lett.} \textbf{\bibinfo{volume}{97}},
  \bibinfo{pages}{223109} (\bibinfo{year}{2010}).

\bibitem[{\citenamefont{De~Padova et~al.}(2010)\citenamefont{De~Padova,
  Quaresima, Ottaviani, Sheverdyaeva, Moras, Carbone, Topwal, Olivieri, Kara,
  Oughaddou et~al.}}]{de_padova_evidence_2010}
\bibinfo{author}{\bibfnamefont{P.}~\bibnamefont{De~Padova}},
  \bibinfo{author}{\bibfnamefont{C.}~\bibnamefont{Quaresima}},
  \bibinfo{author}{\bibfnamefont{C.}~\bibnamefont{Ottaviani}},
  \bibinfo{author}{\bibfnamefont{P.~M.} \bibnamefont{Sheverdyaeva}},
  \bibinfo{author}{\bibfnamefont{P.}~\bibnamefont{Moras}},
  \bibinfo{author}{\bibfnamefont{C.}~\bibnamefont{Carbone}},
  \bibinfo{author}{\bibfnamefont{D.}~\bibnamefont{Topwal}},
  \bibinfo{author}{\bibfnamefont{B.}~\bibnamefont{Olivieri}},
  \bibinfo{author}{\bibfnamefont{A.}~\bibnamefont{Kara}},
  \bibinfo{author}{\bibfnamefont{H.}~\bibnamefont{Oughaddou}},
  \bibnamefont{et~al.}, \bibinfo{journal}{Appl. Phys. Lett.}
  \textbf{\bibinfo{volume}{96}}, \bibinfo{pages}{261905}
  (\bibinfo{year}{2010}).

\bibitem[{\citenamefont{Cahangirov et~al.}(2009)\citenamefont{Cahangirov,
  Topsakal, Akt$\ddot{u}$rk, Scedilahin, and Ciraci}}]{cahangirov_two-_2009}
\bibinfo{author}{\bibfnamefont{S.}~\bibnamefont{Cahangirov}},
  \bibinfo{author}{\bibfnamefont{M.}~\bibnamefont{Topsakal}},
  \bibinfo{author}{\bibfnamefont{E.}~\bibnamefont{Akt$\ddot{u}$rk}},
  \bibinfo{author}{\bibfnamefont{H.}~\bibnamefont{Scedilahin}},
  \bibnamefont{and} \bibinfo{author}{\bibfnamefont{S.}~\bibnamefont{Ciraci}},
  \bibinfo{journal}{Phys. Rev. Lett.} \textbf{\bibinfo{volume}{102}},
  \bibinfo{pages}{236804} (\bibinfo{year}{2009}).

\bibitem[{\citenamefont{Ding and Ni}(2009)}]{ding_electronic_2009}
\bibinfo{author}{\bibfnamefont{Y.}~\bibnamefont{Ding}} \bibnamefont{and}
  \bibinfo{author}{\bibfnamefont{J.}~\bibnamefont{Ni}}, \bibinfo{journal}{Appl.
  Phys. Lett.} \textbf{\bibinfo{volume}{95}}, \bibinfo{pages}{083115}
  (\bibinfo{year}{2009}).

\bibitem[{\citenamefont{Liu et~al.}(2011)\citenamefont{Liu, Feng, and
  Yao}}]{liu_quantum_2011}
\bibinfo{author}{\bibfnamefont{C.}~\bibnamefont{Liu}},
  \bibinfo{author}{\bibfnamefont{W.}~\bibnamefont{Feng}}, \bibnamefont{and}
  \bibinfo{author}{\bibfnamefont{Y.}~\bibnamefont{Yao}},
  \bibinfo{journal}{Phys. Rev. Lett.} \textbf{\bibinfo{volume}{107}},
  \bibinfo{pages}{076802} (\bibinfo{year}{2011}).

\bibitem[{\citenamefont{{Guzm\'an-Verri} and Lew
  Yan~Voon}(2007)}]{guzman-verri_electronic_2007}
\bibinfo{author}{\bibfnamefont{G.~G.} \bibnamefont{{Guzm\'an-Verri}}}
  \bibnamefont{and} \bibinfo{author}{\bibfnamefont{L.~C.} \bibnamefont{Lew
  Yan~Voon}}, \bibinfo{journal}{Phys. Rev. B} \textbf{\bibinfo{volume}{76}},
  \bibinfo{pages}{075131} (\bibinfo{year}{2007}).

\bibitem[{\citenamefont{Hasan and Kane}(2010)}]{hasan_colloquium:_2010}
\bibinfo{author}{\bibfnamefont{M.~Z.} \bibnamefont{Hasan}} \bibnamefont{and}
  \bibinfo{author}{\bibfnamefont{C.~L.} \bibnamefont{Kane}},
  \bibinfo{journal}{Rev. Mod. Phys.} \textbf{\bibinfo{volume}{82}},
  \bibinfo{pages}{3045} (\bibinfo{year}{2010}).

\bibitem[{\citenamefont{Qi and Zhang}(2010{\natexlab{a}})}]{qi_quantum_2010}
\bibinfo{author}{\bibfnamefont{X.}~\bibnamefont{Qi}} \bibnamefont{and}
  \bibinfo{author}{\bibfnamefont{S.}~\bibnamefont{Zhang}},
  \bibinfo{journal}{Physics Today} \textbf{\bibinfo{volume}{63}},
  \bibinfo{pages}{33} (\bibinfo{year}{2010}{\natexlab{a}}).

\bibitem[{\citenamefont{Qi and
  Zhang}(2010{\natexlab{b}})}]{qi_topological_2010}
\bibinfo{author}{\bibfnamefont{X.}~\bibnamefont{Qi}} \bibnamefont{and}
  \bibinfo{author}{\bibfnamefont{S.}~\bibnamefont{Zhang}},
  \bibinfo{journal}{1008.2026}  (\bibinfo{year}{2010}{\natexlab{b}}),
  \urlprefix\url{http://arxiv.org/abs/1008.2026}.

\bibitem[{\citenamefont{Kane and Mele}(2005{\natexlab{a}})}]{kane_quantum_2005}
\bibinfo{author}{\bibfnamefont{C.~L.} \bibnamefont{Kane}} \bibnamefont{and}
  \bibinfo{author}{\bibfnamefont{E.~J.} \bibnamefont{Mele}},
  \bibinfo{journal}{Phys. Rev. Lett.} \textbf{\bibinfo{volume}{95}},
  \bibinfo{pages}{226801} (\bibinfo{year}{2005}{\natexlab{a}}).

\bibitem[{\citenamefont{Yao et~al.}(2007)\citenamefont{Yao, Ye, Qi, Zhang, and
  Fang}}]{yao_spin-orbit_2007}
\bibinfo{author}{\bibfnamefont{Y.}~\bibnamefont{Yao}},
  \bibinfo{author}{\bibfnamefont{F.}~\bibnamefont{Ye}},
  \bibinfo{author}{\bibfnamefont{X.}~\bibnamefont{Qi}},
  \bibinfo{author}{\bibfnamefont{S.}~\bibnamefont{Zhang}}, \bibnamefont{and}
  \bibinfo{author}{\bibfnamefont{Z.}~\bibnamefont{Fang}},
  \bibinfo{journal}{Phys. Rev. B} \textbf{\bibinfo{volume}{75}},
  \bibinfo{pages}{041401} (\bibinfo{year}{2007}).

\bibitem[{\citenamefont{Min et~al.}(2006)\citenamefont{Min, Hill, Sinitsyn,
  Sahu, Kleinman, and {MacDonald}}}]{min_intrinsic_2006}
\bibinfo{author}{\bibfnamefont{H.}~\bibnamefont{Min}},
  \bibinfo{author}{\bibfnamefont{J.~E.} \bibnamefont{Hill}},
  \bibinfo{author}{\bibfnamefont{N.~A.} \bibnamefont{Sinitsyn}},
  \bibinfo{author}{\bibfnamefont{B.~R.} \bibnamefont{Sahu}},
  \bibinfo{author}{\bibfnamefont{L.}~\bibnamefont{Kleinman}}, \bibnamefont{and}
  \bibinfo{author}{\bibfnamefont{A.~H.} \bibnamefont{{MacDonald}}},
  \bibinfo{journal}{Phys. Rev. B} \textbf{\bibinfo{volume}{74}},
  \bibinfo{pages}{165310} (\bibinfo{year}{2006}).

\bibitem[{\citenamefont{Bernevig et~al.}(2006)\citenamefont{Bernevig, Hughes,
  and Zhang}}]{bernevig_quantum_2006}
\bibinfo{author}{\bibfnamefont{B.~A.} \bibnamefont{Bernevig}},
  \bibinfo{author}{\bibfnamefont{T.~L.} \bibnamefont{Hughes}},
  \bibnamefont{and} \bibinfo{author}{\bibfnamefont{S.}~\bibnamefont{Zhang}},
  \bibinfo{journal}{Science} \textbf{\bibinfo{volume}{314}},
  \bibinfo{pages}{1757 } (\bibinfo{year}{2006}).

\bibitem[{\citenamefont{K$\ddot{o}$nig
  et~al.}(2007)\citenamefont{K$\ddot{o}$nig, Wiedmann, Br$\ddot{u}$ne, Roth,
  Buhmann, Molenkamp, Qi, and Zhang}}]{koenig_quantum_2007}
\bibinfo{author}{\bibfnamefont{M.}~\bibnamefont{K$\ddot{o}$nig}},
  \bibinfo{author}{\bibfnamefont{S.}~\bibnamefont{Wiedmann}},
  \bibinfo{author}{\bibfnamefont{C.}~\bibnamefont{Br$\ddot{u}$ne}},
  \bibinfo{author}{\bibfnamefont{A.}~\bibnamefont{Roth}},
  \bibinfo{author}{\bibfnamefont{H.}~\bibnamefont{Buhmann}},
  \bibinfo{author}{\bibfnamefont{L.~W.} \bibnamefont{Molenkamp}},
  \bibinfo{author}{\bibfnamefont{X.}~\bibnamefont{Qi}}, \bibnamefont{and}
  \bibinfo{author}{\bibfnamefont{S.}~\bibnamefont{Zhang}},
  \bibinfo{journal}{Science} \textbf{\bibinfo{volume}{318}},
  \bibinfo{pages}{766 } (\bibinfo{year}{2007}).

\bibitem[{\citenamefont{Murakami}(2006)}]{murakami_quantum_2006}
\bibinfo{author}{\bibfnamefont{S.}~\bibnamefont{Murakami}},
  \bibinfo{journal}{Phys. Rev. Lett.} \textbf{\bibinfo{volume}{97}},
  \bibinfo{pages}{236805} (\bibinfo{year}{2006}).

\bibitem[{\citenamefont{Weeks et~al.}(2011)\citenamefont{Weeks, Hu, Alicea,
  Franz, and Wu}}]{weeks_engineering_2011}
\bibinfo{author}{\bibfnamefont{C.}~\bibnamefont{Weeks}},
  \bibinfo{author}{\bibfnamefont{J.}~\bibnamefont{Hu}},
  \bibinfo{author}{\bibfnamefont{J.}~\bibnamefont{Alicea}},
  \bibinfo{author}{\bibfnamefont{M.}~\bibnamefont{Franz}}, \bibnamefont{and}
  \bibinfo{author}{\bibfnamefont{R.}~\bibnamefont{Wu}},
  \bibinfo{journal}{1104.3282}  (\bibinfo{year}{2011}),
  \urlprefix\url{http://arxiv.org/abs/1104.3282}.

\bibitem[{\citenamefont{Liu et~al.}(2008)\citenamefont{Liu, Hughes, Qi, Wang,
  and Zhang}}]{liu_quantum_2008}
\bibinfo{author}{\bibfnamefont{C.}~\bibnamefont{Liu}},
  \bibinfo{author}{\bibfnamefont{T.~L.} \bibnamefont{Hughes}},
  \bibinfo{author}{\bibfnamefont{X.}~\bibnamefont{Qi}},
  \bibinfo{author}{\bibfnamefont{K.}~\bibnamefont{Wang}}, \bibnamefont{and}
  \bibinfo{author}{\bibfnamefont{S.}~\bibnamefont{Zhang}},
  \bibinfo{journal}{Phys. Rev. Lett.} \textbf{\bibinfo{volume}{100}},
  \bibinfo{pages}{236601} (\bibinfo{year}{2008}).

\bibitem[{\citenamefont{Knez et~al.}(2011)\citenamefont{Knez, Du, and
  Sullivan}}]{knez_evidence_2011}
\bibinfo{author}{\bibfnamefont{I.}~\bibnamefont{Knez}},
  \bibinfo{author}{\bibfnamefont{R.}~\bibnamefont{Du}}, \bibnamefont{and}
  \bibinfo{author}{\bibfnamefont{G.}~\bibnamefont{Sullivan}},
  \bibinfo{journal}{1105.0137}  (\bibinfo{year}{2011}),
  \urlprefix\url{http://arxiv.org/abs/1105.0137}.

\bibitem[{\citenamefont{Kane and Mele}(2005{\natexlab{b}})}]{kane_z2_2005}
\bibinfo{author}{\bibfnamefont{C.~L.} \bibnamefont{Kane}} \bibnamefont{and}
  \bibinfo{author}{\bibfnamefont{E.~J.} \bibnamefont{Mele}},
  \bibinfo{journal}{Phys. Rev. Lett.} \textbf{\bibinfo{volume}{95}},
  \bibinfo{pages}{146802} (\bibinfo{year}{2005}{\natexlab{b}}).

\bibitem[{\citenamefont{Geim and Novoselov}(2007)}]{geim_rise_2007}
\bibinfo{author}{\bibfnamefont{A.~K.} \bibnamefont{Geim}} \bibnamefont{and}
  \bibinfo{author}{\bibfnamefont{K.~S.} \bibnamefont{Novoselov}},
  \bibinfo{journal}{Nature Mater.} \textbf{\bibinfo{volume}{6}},
  \bibinfo{pages}{183} (\bibinfo{year}{2007}).

\bibitem[{\citenamefont{Castro~Neto et~al.}(2009)\citenamefont{Castro~Neto,
  Guinea, Peres, Novoselov, and Geim}}]{castro_neto_electronic_2009}
\bibinfo{author}{\bibfnamefont{A.~H.} \bibnamefont{Castro~Neto}},
  \bibinfo{author}{\bibfnamefont{F.}~\bibnamefont{Guinea}},
  \bibinfo{author}{\bibfnamefont{N.~M.~R.} \bibnamefont{Peres}},
  \bibinfo{author}{\bibfnamefont{K.~S.} \bibnamefont{Novoselov}},
  \bibnamefont{and} \bibinfo{author}{\bibfnamefont{A.~K.} \bibnamefont{Geim}},
  \bibinfo{journal}{Rev. Mod. Phys.} \textbf{\bibinfo{volume}{81}},
  \bibinfo{pages}{109} (\bibinfo{year}{2009}).

\bibitem[{\citenamefont{{Huertas-Hernando}
  et~al.}(2006)\citenamefont{{Huertas-Hernando}, Guinea, and
  Brataas}}]{huertas-hernando_spin-orbit_2006}
\bibinfo{author}{\bibfnamefont{D.}~\bibnamefont{{Huertas-Hernando}}},
  \bibinfo{author}{\bibfnamefont{F.}~\bibnamefont{Guinea}}, \bibnamefont{and}
  \bibinfo{author}{\bibfnamefont{A.}~\bibnamefont{Brataas}},
  \bibinfo{journal}{Phys. Rev. B} \textbf{\bibinfo{volume}{74}},
  \bibinfo{pages}{155426} (\bibinfo{year}{2006}).

\bibitem[{\citenamefont{Slater and Koster}(1954)}]{slater_simplified_1954}
\bibinfo{author}{\bibfnamefont{J.~C.} \bibnamefont{Slater}} \bibnamefont{and}
  \bibinfo{author}{\bibfnamefont{G.~F.} \bibnamefont{Koster}},
  \bibinfo{journal}{Phys. Rev.} \textbf{\bibinfo{volume}{94}},
  \bibinfo{pages}{1498} (\bibinfo{year}{1954}).

\bibitem[{\citenamefont{Saito et~al.}(1992)\citenamefont{Saito, Fujita,
  Dresselhaus, and Dresselhaus}}]{saito_electronic_1992}
\bibinfo{author}{\bibfnamefont{R.}~\bibnamefont{Saito}},
  \bibinfo{author}{\bibfnamefont{M.}~\bibnamefont{Fujita}},
  \bibinfo{author}{\bibfnamefont{G.}~\bibnamefont{Dresselhaus}},
  \bibnamefont{and} \bibinfo{author}{\bibfnamefont{M.~S.}
  \bibnamefont{Dresselhaus}}, \bibinfo{journal}{Phys. Rev. B}
  \textbf{\bibinfo{volume}{46}}, \bibinfo{pages}{1804} (\bibinfo{year}{1992}).

\bibitem[{\citenamefont{Harrison}(1989)}]{harrison_electronic_1989}
\bibinfo{author}{\bibfnamefont{W.~A.} \bibnamefont{Harrison}},
  \emph{\bibinfo{title}{Electronic Structure and the Properties of Solids: The
  Physics of the Chemical Bond}} (\bibinfo{publisher}{Dover Publications},
  \bibinfo{year}{1989}), ISBN \bibinfo{isbn}{0486660214}.

\bibitem[{\citenamefont{Pedersen et~al.}(2010)\citenamefont{Pedersen, Fisker,
  and Jensen}}]{pedersen_tight-binding_2010}
\bibinfo{author}{\bibfnamefont{T.~G.} \bibnamefont{Pedersen}},
  \bibinfo{author}{\bibfnamefont{C.}~\bibnamefont{Fisker}}, \bibnamefont{and}
  \bibinfo{author}{\bibfnamefont{R.~V.} \bibnamefont{Jensen}},
  \bibinfo{journal}{Journal of Physics and Chemistry of Solids}
  \textbf{\bibinfo{volume}{71}}, \bibinfo{pages}{18} (\bibinfo{year}{2010}).

\bibitem[{\citenamefont{Chadi}(1977)}]{PhysRevB.16.790}
\bibinfo{author}{\bibfnamefont{D.~J.} \bibnamefont{Chadi}},
  \bibinfo{journal}{Phys. Rev. B} \textbf{\bibinfo{volume}{16}},
  \bibinfo{pages}{790} (\bibinfo{year}{1977}).

\bibitem[{\citenamefont{Liu et~al.}(2010)\citenamefont{Liu, Qi, Zhang, Dai,
  Fang, and Zhang}}]{liu_model_2010}
\bibinfo{author}{\bibfnamefont{C.}~\bibnamefont{Liu}},
  \bibinfo{author}{\bibfnamefont{X.}~\bibnamefont{Qi}},
  \bibinfo{author}{\bibfnamefont{H.}~\bibnamefont{Zhang}},
  \bibinfo{author}{\bibfnamefont{X.}~\bibnamefont{Dai}},
  \bibinfo{author}{\bibfnamefont{Z.}~\bibnamefont{Fang}}, \bibnamefont{and}
  \bibinfo{author}{\bibfnamefont{S.}~\bibnamefont{Zhang}},
  \bibinfo{journal}{Phys. Rev. B} \textbf{\bibinfo{volume}{82}},
  \bibinfo{pages}{045122} (\bibinfo{year}{2010}).

\bibitem[{\citenamefont{Winkler}(2003)}]{winkler_spin-orbit_2003}
\bibinfo{author}{\bibfnamefont{R.}~\bibnamefont{Winkler}},
  \emph{\bibinfo{title}{Spin-orbit Coupling Effects in {Two-Dimensional}
  Electron and Hole Systems}} (\bibinfo{publisher}{Springer},
  \bibinfo{year}{2003}), \bibinfo{edition}{1st} ed., ISBN
  \bibinfo{isbn}{9783540011873}.

\end{thebibliography}

\end{document}